\documentclass[aps,prc,twocolumn,floatfix,nofootinbib,showpacs,superscriptaddress]{revtex4-1}

\usepackage{amsmath,amsfonts,amssymb,bm}
\usepackage{graphicx}
\usepackage{multirow}
\usepackage{color}
\usepackage{pzccal}
\usepackage{soul}
\usepackage{float}
\usepackage{mathrsfs}

\begin{document}

\newcommand{\dslash}[1]{#1 \hspace{-1.2ex}\slash}

\title{Baryon and meson screening masses}

\author{Kun-lun Wang}
\affiliation{Department of Physics, Center for High Energy Physics and State Key Laboratory of Nuclear Physics and Technology, Peking University, Beijing 100871, China}

\author{Yu-xin Liu}
\email[Corresponding author: ]{yxliu@pku.edu.cn}
\affiliation{Department of Physics, Center for High Energy Physics and State Key Laboratory of Nuclear Physics and Technology, Peking University, Beijing 100871, China}

\author{Lei Chang}
\affiliation{Institut f\"ur Kernphysik, Forschungszentrum J\"ulich, D-52425 J\"ulich, Germany}

\author{Craig D. Roberts}
\email[Corresponding author: ]{cdroberts@anl.gov}
\affiliation{Physics Division, Argonne National Laboratory, Argonne, Illinois 60439, USA}
\affiliation{Department of Physics, Illinois Institute of
Technology, Chicago, Illinois 60616-3793, USA}

\author{Sebastian M.~Schmidt}
\affiliation{Institute for Advanced Simulation, Forschungszentrum J\"ulich and JARA, D-52425 J\"ulich, Germany}

\date{23 January 2013}

\begin{abstract}
In a strongly-coupled quark-gluon plasma, collective excitations of gluons and quarks should dominate over the excitation of individual quasi-free gluon and quark modes.  To explore this possibility, we computed screening masses for ground-state light-quark mesons and baryons at leading-order in a symmetry-preserving truncation scheme for the Dyson-Schwinger equations using a confining formulation of a contact-interaction at nonzero temperature.  Meson screening masses are obtained from Bethe-Salpeter equations; and baryon analogues from a novel construction of the Faddeev equation, which employs an improved quark-exchange approximation in the kernel.  Our treatment implements a deconfinement transition that is coincident with chiral symmetry restoration in the chiral limit, when both transitions are second order.  Despite deconfinement, in all $T=0$ bound-state channels, strong correlations persist above the critical temperature, $T > T_c$; and, in the spectrum defined by the associated screening masses, degeneracy between parity-partner correlations is apparent for $T \gtrsim 1.3\,T_c$.  Notwithstanding these results, there are reasons  (including Golberger-Treiman relations) to suppose that the inertial masses of light-quark bound-states, when they may be defined, vanish at the deconfinement temperature; and that this is a signal of bound-state dissolution.  Where a sensible comparison is possible, our predictions are consistent with results from contemporary numerical simulations of lattice-regularised QCD.
\end{abstract}


\maketitle

\section{Introduction}
It is generally held \cite{national2012Nuclear} that a strongly-coupled quark-gluon plasma (sQGP) has been produced at the relativistic heavy ion collider (RHIC) and that this substance behaves as a nearly perfect fluid on a domain of temperature that extends above that required for its creation: $T_c<T \lesssim 2 \,T_c$ \cite{Qin:2010pc}.  If such is the case, then collective excitations of gluons and quarks should dominate within this domain over the excitation of individual quasi-free gluon and quark modes.  It has long been known that one may explore this possibility by studying the screening masses of hadrons above $T_c$ \cite{DeTar:1987xb}, since the long-range structure of a plasma determines quantities such as its equation of state and transport properties.  It is noteworthy, however, that sufficiently removed from $T=0$, the screening masses in a given channel do not have a simple connection with the inertial masses of what were the channel's bound-states at $T=0$.

In the calculation of masses (screening or inertial), as with most other applications, in order to arrive at robust conclusions, one should employ an approach that preserves the symmetries of QCD, is simultaneously applicable to both mesons and baryons, and has already been applied with success to a diverse array of observables.  The numerical simulation of lattice-regularised QCD is one such approach, from which some results, relevant herein, are described, e.g., in Refs.\,\cite{Pushkina:2004wa,Cheng:2010fe,Petreczky:2012rq}.  It is found that correlations persist in hadron channels above $T_c$ and that parity-partner screening masses become equal.

%
The temperature dependence of hadron masses may also be explored in the continuum using models with an arguable foundation in QCD.  Unfortunately, few such models are simultaneously applicable to studies of mesons and baryons; and amongst those that are, different models and treatments lead to different conclusions.

For example, in a formulation of the Global Colour-symmetry Model \cite{Cahill:1985mh} that treats mesons and baryons differently, it is found \cite{Mo:2010zza} that the nucleon's mass vanishes and the associated radius diverges at the chiral symmetry restoration temperature ($T_c$).
Alternatively, a scalar-diquark-only Faddeev equation treatment of the nucleon,
within a Nambu--Jona-Lasinio model framework \cite{Nambu:1961tp} that does not express confinement, indicates \cite{Mu:2012zz} that whilst the diquark correlation can survive above the temperature for chiral symmetry restoration, the nucleon does not: its mass falls with increasing temperature and, below $T_c$, it lies above the quark$+$scalar-diquark breakup threshold.
One might nevertheless argue that there is some qualitative similarity between these outcomes.

On the other hand, an analysis of the nucleon using thermal finite-energy sum rules produces \cite{Dominguez:1992vg} a nucleon mass that rises rapidly in the neighbourhood of $T_c$, owing, it is suggested, to a vanishing probability for three quarks to form a correlation above this temperature.  In this framework, results for mesons are similar \cite{Dominguez:2012bs}.

These results are all interpreted as statements about inertial masses; and deconfinement is claimed in each framework, although the signal is markedly different.  Plainly, therefore, results obtained in the computation of inertial masses at $\pi T \gtrsim \Lambda_{\rm QCD}$ are very sensitive to assumptions; and such discordant outcomes are unsatisfactory.  This prompts us to re-examine the evolution of hadron masses with temperature in the continuum.  We will, however, focus primarily on the computation of screening masses.  These static quantities are only indirectly sensitive to the real-time scattering processes that may complicate the evaluation of inertial masses, which, in-medium, are all likely to become complex numbers: $M \to m(T) -i \omega(T)$ \cite{Leutwyler:1990uq}.  Finally, however, and in principle, upon computation of all screening masses in a given channel, the inertial masses may be reconstructed.

We choose to employ QCD's Dyson-Schwinger equations (DSEs) in our analysis.  The DSEs have long been used to explore the response of hadron phenomena to nonzero temperature and density \cite{Roberts:2000aa,Hayano:2008vn}.  They are an excellent tool for use in our study because they provide a Poincar\'e-covariant framework that \cite{Maris:2003vk,Chang:2011vu,Bashir:2012fs}: is capable of simultaneously implementing light-quark confinement and expressing dynamical chiral symmetry breaking (DCSB); admits a symmetry-preserving truncation scheme; and provides a unified and uniform treatment of mesons and baryons

Here it is worth commenting further on the manner by which the study of mesons and baryons is unified.  Both are treated as continuum bound-state problems with kernels built from dressed-gluon and dressed-quark propagators that express the dynamical generation of mass: mesons via Bethe-Salpeter equations (BSEs) and baryons via Faddeev equations.  In this way, they are completely equivalent.  Formulation of the Faddeev equations is simplified by capitalising on the established importance of diquark correlations \cite{Cahill:1987qr,Cahill:1988dx}.  This is not unusual.  However, in extending the Faddeev equations to nonzero temperature herein, we retain axial-vector diquark correlations.  They are crucial at zero temperature \cite{Roberts:2011cf,Hecht:2002ej}; e.g., they provide significant attraction in the nucleon bound-state and the $\Delta$-resonance is inaccessible without them.  There is no reason to expect axial-vector diquarks to be less important at $T\neq 0$; and their omission in earlier studies undermines the reliability of those analyses.

In employing the DSE approach herein, we will exploit a symmetry-preserving treatment of a vector$\times$vector contact interaction because it produces $T=0$ results for hadron static properties that are not realistically distinguishable from those obtained by using more sophisticated kernels.  This is demonstrated in Refs.\,\cite{GutierrezGuerrero:2010md,Roberts:2010rn,Roberts:2011cf,Roberts:2011wy,%
Wilson:2011aa,Chen:2012qr,Chen:2012tx}.

In Sec.\,\ref{sec:background} we explain our confining, symmetry-preserving treatment of the contact interaction and its extension to nonzero temperature.  This Section covers the gap, Bethe-Salpeter and Faddeev equations.  It is augmented by two appendices, which detail the derivation of the Faddeev equations.  We highlight at the outset that our Faddeev equations do not express the full complexity that arises in-medium.  Notwithstanding this, they are practical simplifications which should at least yield qualitatively and semi-quantitatively reliable insight into the behaviour of nucleon and $\Delta$ screening masses.

Our formulation of the contact interaction implements confinement, following the notions of Ref.\,\cite{Ebert:1996vx}.  This distinguishes it from most previous simultaneous, covariant, continuum treatments of mesons and baryons in-medium.  On the other hand, we expect QCD to exhibit deconfinement at some $T=T_d>0$.  Consequently, a dynamical mechanism should be incorporated the allows for deconfinement.  In a related context, this was considered in Ref.\,\cite{Blaschke:2001yj}.  Our different approach is described in Sec.\,\ref{bag}.

Our results are described in Sec.\,\ref{numerical}.  They range from an analysis and illustration of chiral symmetry restoration and deconfinement to a description of the $T$-dependence of screening masses of ground-state light-quark mesons, diquark correlations and baryons.  A key result is that correlations typically persist in hadron channels above the temperature associated with chiral symmetry restoration and deconfinement.  As we explain, however, this is not a statement that true bound-states persist.

In Sec.\,\ref{sec:GTR}, following Ref.\,\cite{Chang:2012cc}, we explore the implications of chiral-symmetry restoration and deconfinement for quark- and nucleon-level Goldberger-Treiman relations; and therefrom, the nucleon's inertial mass.  Section~\ref{summary} presents a summary and perspective.

\section{Contact interaction at nonzero temperature}
\label{sec:background}
The formulation of DSEs at nonzero temperature is described in Refs.\,\cite{Roberts:2000aa,Bashir:2012fs} so here we proceed directly with a specific discussion of the contact interaction described in App.\,\ref{sec:contact}.

\subsection{Dressed quark propagator}
\label{sec:Sp}
The $T\neq 0$ dressed-quark propagator is obtained from the following gap equation:
\begin{eqnarray}
\nonumber S^{-1}(\vec p,\omega_n) &=&
i\vec\gamma\cdot \vec p + i\gamma_{4}^{} \omega_{n}^{} + m  \\
&&  + \frac{16\pi\alpha_{\rm IR}}{3 m_{G}^{2}}  \int_{l,dq}
\gamma_{\mu}^{} S(\vec q,\omega_{l}^{} )
\gamma_{\mu}^{} \, ,
\label{Rgap}
\end{eqnarray}
where $m=m_u=m_d$ is the light-quark current-mass, $\int_{l,dq} = T \sum_{l=-\infty}^\infty \int d^3\vec q/(2\pi)^3$ and $\omega_n=(2n+1)\pi T$ is the fermion Matsubara frequency.  Equation\,\eqref{Rgap} is the rainbow approximation to the gap equation, which is the leading term in the systematic, symmetry-preserving truncation scheme of Ref.\,\cite{Munczek:1994zz,Bender:1996bb}.

The solution of Eq.\,\eqref{Rgap} is momentum-independent, as it was for $T=0$; viz.,
\begin{equation}
\label{SpT}
S^{-1}(\vec p,\omega_n) =
i\vec\gamma\cdot \vec p + i\gamma_{4}^{} \omega_{n}^{} + M\,,
\end{equation}
where the dressed-quark mass is determined from
\begin{eqnarray}
M & = & m +
\frac{16\pi\alpha_{\rm IR}}{3 m_{G}^{2}}  \int_{l,dq}
\frac{4 M}{s_l + M^{2}}\,, \label{eq:gap}
\end{eqnarray}
with $s_l = \vec{q}^{\,2} + \omega_{l}^{2}$.  The integral is linearly divergent and may be regularised using the $T=0$ procedure; namely, we write \cite{Ebert:1996vx}
\begin{eqnarray}
\nonumber
\frac{1}{s_l+M^2} & = & \int_0^\infty d\tau\,{\rm e}^{-\tau (s_l+M^2)}  \\
&\rightarrow&  \int_{\tau_{\rm uv}^2}^{\tau_{\rm ir}^2} d\tau\,{\rm e}^{-\tau (s_l+M^2)}
\label{RegC}\\
 &=&
\frac{{\rm e}^{- (s_l+M^2)\tau_{\rm uv}^2}-{\rm e}^{-(s_l+M^2) \tau_{\rm ir}^2}}{s_l+M^2} \,, \label{eq:regulator}
\end{eqnarray}
where $\tau_{\rm ir,uv}$ are, respectively, infrared and ultraviolet regulators.  Since the interaction in Eq.\,(\ref{njlgluon}) does not define a renormalisable theory, then $\Lambda_{\rm uv}:=1/\tau_{\rm uv}$ cannot be removed but instead plays a dynamical role, setting the scale of all dimensioned quantities.  Using Eq.\,\eqref{eq:regulator}, Eq.\,\eqref{eq:gap} becomes
\begin{equation}
M = m + M\frac{4\alpha_{\rm IR}}{3\pi m_G^2}\,\,{\cal C}^{\rm iu}(M^2;T)\,,
\label{gapactual}
\end{equation}
where ${\cal C}^{\rm iu}(M^2;T)$ is defined in Eq.\,\eqref{CirT} and the parameters in Table~\ref{Tab:DressedQuarks}.

It is apparent from the rightmost expression in Eq.\,(\ref{eq:regulator}) that a finite value of $\tau_{\rm ir}=1/\Lambda_{\rm ir}$ implements confinement by ensuring the absence of quark production thresholds in all processes \cite{Roberts:2000aa,Bashir:2012fs}.\footnote{The potential between infinitely-heavy quarks measured in numerical simulations of quenched lattice-regularised QCD -- the so-called static potential -- is not related in any known manner to the question of confinement in the real world, in which light quarks are ubiquitous.  It is a basic feature of QCD that light-particle creation and annihilation effects are essentially nonperturbative and therefore it is impossible in principle to compute a potential between two light quarks \cite{Bali:2005fu,Chang:2009ae}.  Confinement may, instead, be related to the analytic properties of QCD's propagators and vertices \cite{Gribov:1999ui,Munczek:1983dx,Stingl:1983pt,Cahill:1988zi,Krein:1990sf,%
Dokshitzer:2004ie,Roberts:2007ji}.
}
This is appropriate for studies of $T=0$ phenomena.  However, we expect QCD to exhibit deconfinement at some $T=T_d>0$, whereat the production thresholds reappear, as illustrated in Refs.\,\cite{Bender:1996bm,Bender:1997jf}.  Therefore, we subsequently introduce a dynamical mechanism that makes $\tau_{\rm ir}$ temperature-dependent.

In the gap equation above, and in the bound-state equations to follow, we omit the temperature-induced separation of the gluon propagator dressing into transverse and longitudinal parts.  Instead, we assume the interaction dressing is frozen at its $T=0$ form.  This is a defect that our study shares with all other continuum analyses of bound-state screening masses.  In our judgement, however, it is not a crippling weakness.  Temperature does affect the nature of gluon dressing but, on the domain of concern to us; i.e., $T \lesssim 2 T_c$, these effects are modest \cite{Cucchieri:2007ta,Aouane:2012bk}.

\subsection{Mesons}
\label{sec:mesons}
\subsubsection{Screening masses}
A strength of the DSE framework is its ability to treat mesons and baryons on an equal footing, both at zero and nonzero temperature.  We illustrate and exploit that capacity herein.

At leading order in the symmetry-preserving truncation scheme of Ref.\,\cite{Bender:1996bb}, one considers the homogeneous rainbow-ladder BSE for mesons; namely,
\begin{equation}
\Gamma_H(Q_0) =  - \frac{16 \pi}{3} \frac{\alpha_{\rm IR}}{m_G^2}
 \int_{l,dt} \gamma_\mu S(t+Q_0) \Gamma_H(Q_0)S (t) \gamma_\mu \,,
\label{LBSEI}
\end{equation}
where $S$ is obtained from Eq.\,\eqref{Rgap} and $Q_0=\{\vec{Q},0\}$ is the total momentum entering the amplitude. (For our immediate purposes, it is necessary to focus only on the meson's zeroth Matsubara frequency.)  The rainbow-ladder truncation is known to provide reliable results for the $T=0$ properties of vector and flavour nonsinglet pseudoscalar mesons \cite{Chang:2011vu,Qin:2011xq}.  Equation\,\eqref{LBSEI} is an eigenvalue problem: it has a solution for $Q_0^2=-m_H^2$, where $m_H$ is the mass of any one of the bound-state's in this channel at $T=0$, owing to $O(4)$-invariance, and the associated screening mass for $T>0$.

We reiterate here that, sufficiently removed from $T=0$; namely, for $\pi T \gtrsim M(T=0)$, a particular screening mass in a given channel has no simple connection with the inertial mass of any of the channel's bound-states at $T=0$.  Indeed, a given channel supports many bound-states at $T=0$ and hence there will necessarily be at least as many screening masses.  Moreover, traceable to each $T=0$ bound-state, there is potentially a different screening mass for each one of the enumerable infinity of (meson or baryon) Matsubara modes.  In order to reconstruct a real-time Green function, from which an inertial mass may be determined, one must: calculate each such mass; compute a Fourier transform of the form
\begin{equation}
\label{BeginWightmann}
\hat {\mathcal S}(\tau) = T \! \sum_{n=-\infty}^\infty\, {\rm e}^{- i \nu_n \tau} \mathcal S(\nu_n)\,,
\end{equation}
where $\{\nu_n\}$ are boson or fermion Matsubara frequencies and $\mathcal S$ is a thermal Schwinger function; complete an analytic continuation via $\tau\to
\tau + i t$ and the limit $\tau\to 0^+$; and finally arrange appropriate step-function-weighted combinations of the result.

\subsubsection{Pseudoscalar- and vector-mesons}
It was shown in Ref.\,\cite{Maris:1997hd} that a pseudoscalar meson must possess components in its Bethe-Salpeter amplitude that may be described as pseudovector in character.  These components play a critical role in the $T=0$ physics of pseudoscalar mesons \cite{Maris:1998hc,GutierrezGuerrero:2010md,Roberts:2010rn,Chang:2012cc,Chen:2012tx}.
This means herein that Eq.\,\eqref{LBSEI} supports a solution in the pseudoscalar channel of the form:
\begin{equation}
\label{piBSA}
\Gamma_\pi(Q_0) = i \gamma_5 E_\pi(Q_0) + \frac{1}{M}\gamma_5 \gamma\cdot Q_0 F_\pi(Q_0)\,.
\end{equation}
In the vector channel there are two distinct components at $T\neq 0$:
\begin{equation}
\Gamma_\rho(Q_0) =
\left\{\begin{array}{l}
\gamma_4 \, E_\rho^\parallel(Q_0) \\
\vec{\gamma}_\perp E_\rho^\perp (Q_0)
\end{array}\right.\,,
\end{equation}
where $\vec{\gamma}_\perp = P_{ij}(Q_0) \gamma_j = (\delta_{ij} - Q_i Q_j/|\vec{Q}|^2)\gamma_j$, $i,j=1,2,3$.  In a symmetry preserving treatment of the interaction in Eq.\,\eqref{njlgluon}, there can be no dependence on a relative momentum in either case.

Explicit forms for the BSEs of the $\pi$- and $\rho$-mesons are readily obtained following the procedures described in Refs.\,\cite{Roberts:2011wy,Chen:2012qr}.  That for the pion is an obvious analogue of Eqs.\,(31)--(35) in Ref.\,\cite{Roberts:2011wy}.  Herein, as ${\cal C}^{\rm iu}(\varsigma)$ is replaced by ${\cal C}^{\rm iu}(\varsigma;T)$, so
\begin{equation}
\overline{\cal C}_1^{\rm iu}(\varsigma)
\to \overline{\cal C}_1^{\rm iu}(\varsigma;T) =
-\frac{d}{d\varsigma} {\cal C}^{\rm iu}(\varsigma;T)\,.
\end{equation}
The argument of these functions is given by ($\hat\alpha = 1-\alpha$)
\begin{equation}
\varsigma = \varsigma(M^2,\alpha,Q_0^2)=M^2 + \alpha \hat \alpha Q_0^2\,.
\end{equation}
Note that we follow the interaction definition of Ref.\,\protect\cite{Chen:2012qr}, so that $1/m_G^2$ in Ref.\,\protect\cite{Roberts:2011wy} becomes $4\pi\alpha_{\rm IR}/m_G^2$ herein.  (See Eq.\,(20) in Ref.\,\protect\cite{Chen:2012qr}.)  For brevity, we will sometimes write
\begin{equation}
\frac{1}{\mathpzc{m}_G^2} = \frac{4\pi\alpha_{\rm IR}}{m_G^2} \,.
\end{equation}

Given these observations, one can readily express the pseudoscalar BSE:
\begin{equation}
\label{bsefinal0}
\left[
\begin{array}{c}
E_\pi(Q_0)\\
F_\pi(Q_0)
\end{array}
\right]
= \frac{4\alpha_{\rm IR}}{3 \pi m_G^2}
\left[
\begin{array}{cc}
{\cal K}^\pi_{EE} & {\cal K}^\pi_{EF} \\
{\cal K}^\pi_{FE} & {\cal K}^\pi_{FF}
\end{array}\right]
\left[\begin{array}{c}
E_\pi(Q_0)\\
F_\pi(Q_0)
\end{array}
\right],
\end{equation}
where
\begin{subequations}
\label{pionkernel}
\begin{eqnarray}
\nonumber
{\cal K}^\pi_{EE} &= &\int_0^1d\alpha \left[ {\cal C}^{\rm iu}(\varsigma(M^2,\alpha,-m_\pi^2);T)\right. \\
&& \left.  + 2 \alpha\hat\alpha \, m_\pi^2 \, \overline{\cal C}^{\rm iu}_1(\varsigma(M^2,\alpha,-m_\pi^2);T)\right],\label{KpiEE}\\
{\cal K}^\pi_{EF} &=& -m_\pi^2 \int_0^1d\alpha\, \overline{\cal C}^{\rm iu}_1(\varsigma(M^2,\alpha,-m_\pi^2);T), \label{KpiEF}\\
{\cal K}^\pi_{FE} &=& \frac{1}{2} M^2 \int_0^1d\alpha\, \overline{\cal C}^{\rm iu}_1(\varsigma(M^2,\alpha,-m_\pi^2);T), \label{KpiFE}\\
{\cal K}^\pi_{FF} &=& - 2 {\cal K}_{FE}\,. \label{KpiFF}
\end{eqnarray}
\end{subequations}

In deriving Eqs.\,\eqref{bsefinal0}, \eqref{pionkernel}, one must use the following identity:
\begin{equation}
\label{WGTIT}
0= \int_0^1d\alpha\, \bigg[{\cal C}^{\rm iu}(\varsigma;T)+{\cal C}_1^{\rm iu}(\varsigma;T)
+{\cal R}^{\rm iu}(\varsigma;T)\bigg],
\end{equation}
where ${\cal R}^{\rm iu}(\varsigma;T)$ is defined in Eq.\,\eqref{eq:RsT}.  As the $T\neq 0$ generalisation of Eq.\,(20) in Ref.\,\cite{Roberts:2011wy}, Eq.\,\eqref{WGTIT} is necessary and sufficient to guarantee the vector and axial-vector Ward-Green-Takahashi identities \cite{Ward:1950xp,Green:1953te,Takahashi:1957xn}.  N.B.\ ${\cal R}^{\rm iu}(\varsigma;T\to 0) = 0$, as shown in connection with Eqs.\,\eqref{eq:2TJtheta0}, \eqref{eq:RsT}.

It is necessary to employ the canonically normalised Bethe-Salpeter amplitude in the computation of observables.  For the pion, that amplitude satisfies
\begin{equation}
1 = \left. \frac{d}{dQ_0^2}\Pi_\pi(K,Q_0) \right|_{K=Q_0},
\end{equation}
where
\begin{eqnarray}
\nonumber \lefteqn{\Pi_\pi(K,Q_0) }\\
&=& 6\, {\rm tr}_{\rm D}
\int_{l,dq} \Gamma_\pi(-K) S(\vec{q}+\vec{Q},\omega_l)\Gamma_\pi(K) S(\vec{q},\omega_l)\,. \quad
\end{eqnarray}

The BSE for the $\rho$-meson's longitudinal component is
\begin{equation}
\label{eq:rhoparallel}
0 = 1 +  {\cal K}^{\rho^\parallel}(-m^2_{\rho^\parallel})\,,
\end{equation}
where
\begin{eqnarray}
\nonumber
{\cal K}^{\rho^\parallel}(z) & = &\frac{4 \alpha_{\rm IR}}{3 \pi m_G^2}
\int_0^1\! d\alpha\, \bigg[\alpha\hat \alpha z \overline{\cal C}_1^{\rm iu}(\varsigma(M^2,\alpha,z);T) \\
&&  \quad + {\cal R}^{\rm iu}(\varsigma(M^2,\alpha,z);T)\bigg]\,.
\label{rhoparallel}
\end{eqnarray}
The canonical normalisation condition for the longitudinal amplitude is
\begin{equation}
\label{rhoparallelnorm}
\frac{1}{(E_\rho^\parallel)^2} =
\left. - 9 \mathpzc{m}_G^2 \frac{d}{dz}
{\cal K}^{\rho^\parallel}(z)\right|_{z=-m^2_{\rho^\parallel}}.
\end{equation}

The BSE for the perpendicular component of the $\rho$-meson is almost identical to Eq.\,\eqref{eq:rhoparallel} except that one omits ${\cal R}^{\rm iu}$ in mapping ${\cal K}^{\rho^\parallel} \to {\cal K}^{\rho^\perp}$.  The equation thus obtained is just Eq.\,(36) in Ref.\,\cite{Roberts:2011wy} with the replacement $\overline{\cal C}_1^{\rm iu}(\varsigma)\to \overline{\cal C}_1^{\rm iu}(\varsigma;T)$.  The canonical normalisation condition is simply Eq.\,\eqref{rhoparallelnorm} with $\parallel \to \perp$ in the obvious places.  At zero temperature the equations for the longitudinal and transverse components are naturally identical, an outcome that is realised because ${\cal R}^{\rm iu}(\varsigma;T\to 0)=0$. 

\subsubsection{Scalar and pseudovector channels}
\label{scalarpseudovector}
The large splitting between parity partners is a striking feature of QCD's spectrum.  It is discussed at length in Refs.\,\cite{Chang:2009zb,Chang:2011ei}, which showed that nonperturbative DCSB-corrections to the rainbow-ladder truncation generate a large spin-orbit repulsion and are responsible for the splitting.  Informed by those analyses, Refs.\,\cite{Roberts:2011wy,Chen:2012qr} modified the rainbow-ladder BSEs in the scalar and pseudovector channels, including a single, common coupling parameter $g_{\rm SO}^0$ whose presence simulates the repulsive effect.  The value
\begin{equation}
g_{\rm SO}^{T=0} = 0.24
\end{equation}
reproduces the experimental value for the $a_1$-$\rho$ splitting.  It is noteworthy that the shift in $m_{a_1}$ is accompanied by an increase of $m_\sigma$, the new value of which matches an estimate for the $\bar q q$-component (dressed-quark core) of the $\sigma$-meson obtained using unitarised chiral perturbation theory \cite{Pelaez:2006nj,RuizdeElvira:2010cs}.

We emulate Refs.\,\cite{Roberts:2011wy,Chen:2012qr} by including $g_{\rm SO}$ in our BSEs for the scalar and pseudovector channels.  However, in anticipation that chiral symmetry is restored above some critical temperature, we enable the parameter's strength to track that of DCSB; viz., we use
\begin{equation}
\label{gSOT}
g^2_{\rm SO} \to g^2_{\rm SO}(T) = 1 - \frac{M(T)}{M(0)}(1-[g^{T=0}_{\rm SO}]^2)\,,
\end{equation}
where $M(T)$ is the $T$-dependent dressed-quark mass obtained from Eq.\,\eqref{eq:gap}.

With a symmetry-preserving regularisation of the contact interaction, the scalar meson Bethe-Salpeter amplitude takes the simple form
\begin{equation}
\Gamma_\sigma(Q_0) = {\mathbf I}_{\rm D} \, E_\sigma(Q_0) \,.
\end{equation}
The screening mass is obtained from
\begin{equation}
1=  \frac{4\alpha_{\rm IR}}{3\pi m_G^2} {\cal K}^{\sigma}(-m_\sigma^2)\,,
\end{equation}
where
\begin{eqnarray}
\nonumber
{\cal K}^{\sigma}(z) &=&
g^2_{\rm SO}(T) \int_0^1\! d\alpha\, \big[
{\cal C}^{\rm iu}(\varsigma(M^2,\alpha,z);T) \\
&& \quad - 2 {\cal C}_1^{\rm iu}(\varsigma(M^2,\alpha,z);T)\big]\,;
\label{Ksigma}
\end{eqnarray}
and the amplitude is canonically normalised via
\begin{equation}
\frac{1}{E_\sigma^2} =
\left.  \frac{3}{2\pi^2} \frac{d}{dz}{\cal K}^{\sigma}(z)\right|_{z=-m^2_{\sigma}}.
\end{equation}

Akin to the $\rho$-meson, at nonzero temperature the Bethe-Salpeter amplitude for the $a_1$ channel has the form
\begin{equation}
\Gamma_{a_1}(Q_0) =
\left\{\begin{array}{l}
\gamma_5 \gamma_4 \, E_{a_1}^\parallel(Q_0) \\
\gamma_5 \vec{\gamma}_\perp E_{a_1}^\perp (Q_0)
\end{array}\right.\,.
\end{equation}
The screening mass of the transverse component is obtained from
\begin{equation}
0=1 + {\cal K}_{a_1^\perp}(-m_{a_1^\perp}^2)\,,
\end{equation}
where
\begin{equation}
\label{Ka1perp}
{\cal K}_{a_1^\perp}(z) = g^2_{\rm SO}(T)
\frac{4\alpha_{\rm IR}}{3 \pi m_G^2}
\int_0^1\! d\alpha\, {\cal C}_1^{\rm iu}(\varsigma(M^2,\alpha,z);T)\,.
\end{equation}
The canonical normalisation condition is
\begin{equation}
\frac{1}{E_{a_1^\perp}^2} = - \left. 9 \mathpzc{m}_G^2 \frac{d}{dz}{\cal K}_{a_1^\perp}(z) \right|_{z=-m_{a_1^\perp}^2}.
\end{equation}

The BSE for the longitudinal component is modified similarly to that of the $\rho$, Eq.\,\eqref{rhoparallel}:
\begin{equation}
\label{a1parallel}
0 = 1 + {\cal K}_{a_1^\parallel}(-m_{a_1^\parallel}^2)\,,
\end{equation}
where
\begin{eqnarray}
\nonumber
{\cal K}_{a_1^\parallel}(z) &=& g^2_{\rm SO}(T)
\frac{4\alpha_{\rm IR}}{3 \pi m_G^2}
\int_0^1\! d\alpha\, \big[ {\cal C}_1^{\rm iu}(\varsigma(M^2,\alpha,z);T) \\
&& \quad + {\cal R}^{\rm iu}(\varsigma(M^2,\alpha,z);T)\big]\,;
\label{Ka1parallel}
\end{eqnarray}
and the canonical normalisation condition is
\begin{equation}
\frac{1}{E_{a_1^\parallel}^2} = - \left. 9 \mathpzc{m}_G^2 \frac{d}{dz}{\cal K}_{a_1^\parallel}(z) \right|_{z=-m_{a_1^\parallel}^2}.
\end{equation}

\subsubsection{Vertex residues}
In our subsequent analysis of deconfinement and chiral symmetry restoration we will display the response to changes in temperature of the residues connected with the $\pi$- and $\sigma$-meson screening masses in, respectively, the pseudovector and pseudoscalar vertices, and the scalar vertex.

In the isospin-symmetry limit, which we employ herein, there is no $\sigma$-meson pole in the vector vertex.  On the other hand, the pseudovector vertex does exhibit a pion pole and the residue is the pion's leptonic decay constant.  At $T\neq 0$ the expression for the relevant decay constant may be derived following Refs.\,\cite{Roberts:2011wy,Maris:2000ig}:
\begin{equation}
\label{fpiT}
f_\pi = \frac{1}{M}\frac{3}{2\pi^2}\big[ E_\pi - 2 F_\pi \big]
{\cal K}^\pi_{FE}\,,
\end{equation}
where ${\cal K}^\pi_{FE}$ is given in Eq.\,\eqref{KpiFE}.  This is actually $f_\pi^\perp$. The expression for $f_\pi^\parallel$ is different and of less interest herein.

The pseudoscalar vertex also exhibits a pion pole.  Its residue is \cite{Roberts:2011wy,Maris:2000ig}
\begin{equation}
\label{residuepi}
\mathpzc{r}_\pi =  \frac{3}{4\pi^2}
\big[ E_\pi {\cal K}^\pi_{EE}  + F_\pi {\cal K}^\pi_{EF}\big]\,,
\end{equation}
where the kernels are given in Eqs.\,\eqref{KpiEE}, \eqref{KpiEF}.  The product $\mathpzc{r}_\pi f_\pi$ defines the in-pion condensate \cite{Brodsky:2010xf,Chang:2011mu,Brodsky:2012ku}.

The scalar vertex exhibits a $\sigma$-meson pole.  Its residue may be derived following Refs.\,\cite{Maris:2000ig,Chang:2011mu}:
\begin{equation}
\label{residuesigma}
\mathpzc{r}_\sigma = \frac{3}{4\pi^2} E_\sigma
{\cal K}^{\sigma}(-m_\sigma^2)\,,
\end{equation}
where ${\cal K}^{\sigma}$ is given in Eq.\,\eqref{Ksigma}.  This residue can be used to express the in-$\sigma$-meson condensate \cite{Chang:2011mu}.

\subsection{Diquark correlations}
\label{sec:qqBSA}
The relevance of rainbow-ladder truncation meson-BSEs to baryon Faddeev equations is explained, e.g., in Sect.\,2.1 of Ref.\,\cite{Roberts:2011cf}; namely, in this truncation one may obtain the mass and Bethe-Salpeter amplitude for a colour-antitriplet quark-quark correlation (diquark) with spin-parity $J^P$ from the equation for a $J^{-P}$-meson in which the only change is a halving of the interaction strength \cite{Cahill:1987qr}.  The flipping of the sign in parity occurs because intrinsic parity is opposite for fermions and antifermions.  N.B.\ Only scalar and axial-vector diquark correlations are needed for the ground-state nucleon and $\Delta$ because these correlations have the same parity as those baryons and masses which are lower.

We remark that the rainbow-ladder truncation generates asymptotic diquark states.  Such states are not observed and their appearance is an artefact of the truncation. Higher-order terms in the quark-quark scattering kernel, whose analogue in the quark-antiquark channel do not materially affect the properties of vector and flavour non-singlet pseudoscalar mesons, ensure that QCD's quark-quark scattering matrix does not exhibit singularities which correspond to asymptotic diquark states \cite{Bender:1996bb,Bender:2002as,Bhagwat:2004hn}.   Studies with kernels that exclude diquark bound states nevertheless support a physical interpretation of the masses, $m_{(qq)_{\!J^P}}$, obtained using the rainbow-ladder truncation; viz., the quantity $\ell_{(qq)^{\!J^P}}:=1/m_{(qq)_{\!J^P}}$ may be interpreted as a range over which the diquark correlation can propagate before losing its identity through fragmentation.

Following these observations, it is straightforward to infer the BSEs for diquark correlations from the formulae in Sec.\,\ref{sec:mesons}.  The Bethe-Salpeter amplitude for a spin-parity $J^P$ diquark is equivalent in form to that for a $J^{-P}$ meson; e.g., the $0^+$ diquark is described by an amplitude $\Gamma_{0^+}(Q_0)$ and
\begin{subequations}
\begin{eqnarray}
\Gamma_{0^+}^C(Q_0) &= & \Gamma_{0^+}(Q_0) C^\dagger\\
&=& i \gamma_5 E_{0^+}(Q_0) + \frac{1}{M}\gamma_5 \gamma\cdot Q_0 F_{0^+}(Q_0)\,,\quad
\end{eqnarray}
\end{subequations}
where $C=\gamma_2\gamma_4$ is the charge-conjugation matrix [see Eqs.\,\eqref{chargec}, \eqref{chargecT}].  The amplitude $\Gamma_{0^+}^C(Q_0)$ and the mass of this correlation are obtained from an equation with the same form as that for the pion except for inclusion in the kernel of a multiplicative factor of $1/2$.  (See, e.g., Eq.\,(23) in Ref.\,\cite{Chen:2012qr}.)

As we are concerned with just the nucleon and $\Delta$, only one more BSE is needed; namely, that for the axial-vector diquark.  The Bethe-Salpeter amplitude for this correlation is constructed from
\begin{equation}
\Gamma_{1^+}^C(Q_0) =
\left\{\begin{array}{l}
\gamma_4 \, E_{1_\parallel^+}(Q_0) \\
\vec{\gamma}_\perp E_{1_\perp^+} (Q_0)
\end{array}\right.\,.
\end{equation}
The mass of the longitudinal component of the correlation is the solution of
\begin{equation}
0 = 1 +  \frac{1}{2}{\cal K}^{\rho^\parallel}(-m^2_{1^+_\parallel})\,,
\end{equation}
where ${\cal K}^{\rho^\parallel}$ is given in Eq.\,\eqref{eq:rhoparallel}; and that of the transverse component from the same equation except that one omits ${\cal R}^{\rm iu}$ in mapping ${\cal K}^{\rho^\parallel} \to {\cal K}^{\rho^\perp}$.  The amplitudes are canonically normalised as follows:
\begin{subequations}
\begin{eqnarray}
\frac{1}{E_{1_\parallel^+}^2} &=& \left. -6  \mathpzc{m}_G^2 \frac{d}{dz} {\cal K}^{\rho^\parallel}(z)\right|_{z=-m^2_{1^+_\parallel}} , \\
\frac{1}{E_{1_\perp^+}^2} &=& \left. -6 \mathpzc{m}_G^2 \frac{d}{dz} {\cal K}^{\rho^\perp}(z)\right|_{z=-m^2_{1^+_\perp}} .
\end{eqnarray}
\end{subequations}

\subsection{Baryon Faddeev Equations}
\label{subsec:baryon}
\begin{figure}[t]
\centerline{%
\includegraphics[clip,width=0.45\textwidth]{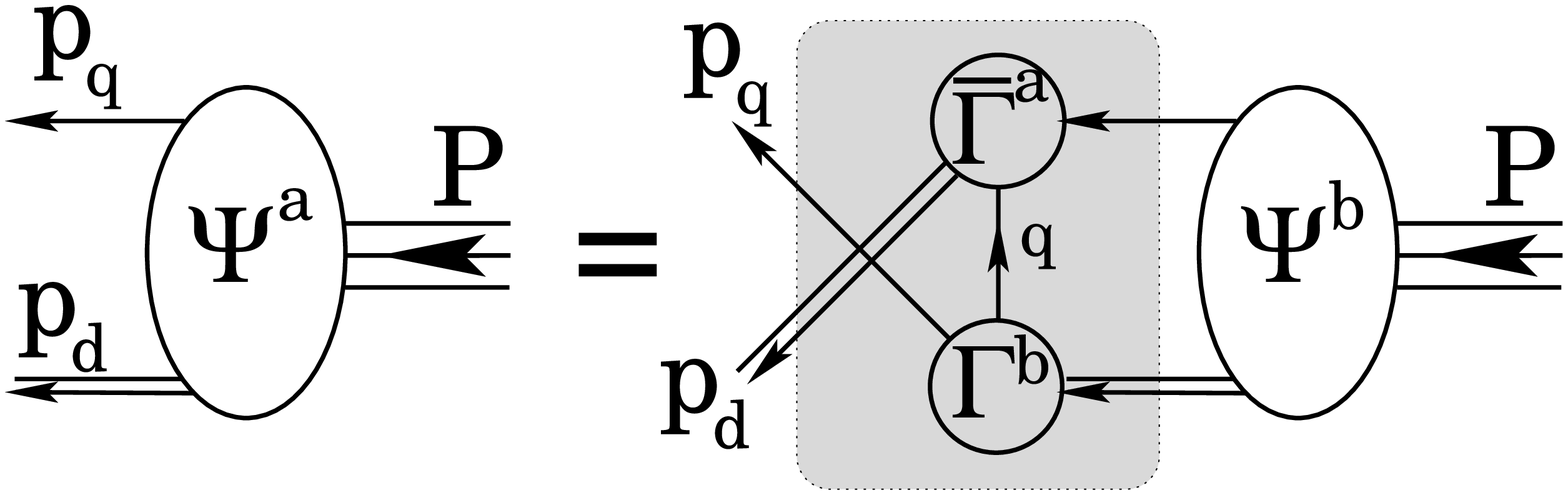}}
\caption{\label{fig:Faddeev} Poincar\'e covariant Faddeev equation, the basis for our computation of baryon screening masses.  $\Psi$ is the Faddeev amplitude for a baryon of total momentum $P= p_q + p_d$.  The shaded region demarcates the kernel of the Faddeev equation, Sec.\,\protect\ref{app:DeltaFE}, in which: the \emph{single line} denotes the dressed-quark propagator, Sec.\,\protect\ref{sec:Sp};
$\Gamma$ is the diquark Bethe-Salpeter amplitude, Sec.\,\protect\ref{sec:qqBSA};
and the \emph{double line} is the diquark propagator, Eqs.\,(\ref{propAVqq}), (\ref{propSCqq}).}
\end{figure}

We base our description of the dressed-quark-core of the nucleon and $\Delta$ resonance on the Faddeev equation introduced in Ref.\,\cite{Cahill:1988dx}, depicted in Fig.\,\ref{fig:Faddeev}, and since studied extensively at $T=0$ (see, e.g., Refs.\,\cite{Hanhart:1995tc,Asami:1995xq,Mineo:1999eq,Hecht:2002ej,Eichmann:2008ef,Cloet:2008re,%
Roberts:2011cf,Eichmann:2011ej,Chen:2012qr}).  The phrase ``dressed-quark-core'' means that, consistent with the rainbow-ladder treatment of mesons and diquark correlations, we deliberately omit contributions to baryon masses that arise from resonant (meson cloud) contributions.  The nature and implications of this omission at $T=0$ are detailed, e.g., in Ref.\,\cite{Roberts:2011cf} (particularly, Sec.\,4.2) and Ref.\,\cite{Chen:2012qr} (particularly, Sec.\,3.3).


With the loss of $O(4)$ invariance at nonzero temperature, a complete description of $J=\frac{1}{2}$, $\frac{3}{2}$ baryons becomes complicated.  In the case of the nucleon, as with the dressed-quark in Eq.\,\eqref{SpT}, the complexity begins with a new structure in the propagator; and then one must account for the separation of the axial-vector diquark into two components.  Notwithstanding this, a careful symmetry-preserving formulation using the contact-interaction could conceivably yield a tractable albeit cluttered problem.  For the $\Delta$-resonance, on the other hand, the complexity becomes extreme \cite{Korpa:2004sh}.

In this first study, therefore, we choose to work with the zero temperature Faddeev equations modified in a simple manner.  Namely, in deriving the Faddeev equations we: ignore $T$-dependent modifications of the propagators and amplitudes; and then realise $T$-dependence in the resulting equations by replacing the $dl_4$ relative-momentum integral by a Matsubara sum, employing the $T$-dependent mass of the dressed-quark and the screening masses of the diquarks (transverse mode, in the case of the axial-vector), and the appropriately matched $T$-dependent diquark Bethe-Salpeter amplitudes.  The procedure is exemplified via the $\Delta$ in App.\,\ref{app:DeltaFE} and the nucleon Faddeev equation is described in App.\,\ref{app:Nucleon}.

\section{Confinement length at nonzero temperature}
\label{bag}
Recall now the rightmost expression in Eq.\,(\ref{eq:regulator}): a finite value of $\tau_{\rm ir}=1/\Lambda_{\rm ir}\approx 0.8\,$fm implements confinement by ensuring the absence of quark production thresholds in all processes \cite{Roberts:2000aa,Bashir:2012fs}.  We expect that QCD exhibits deconfinement at some $T=T_d>0$, whereat the production thresholds reappear, as illustrated in Ref.\,\cite{Bender:1996bm}.  Here we therefore introduce a dynamical mechanism that makes $\tau_{\rm ir}$ temperature-dependent.

Whilst more sophistication is required in general \cite{Qin:2010nq}, if one works in the rainbow-ladder truncation, then chiral symmetry restoration and deconfinement may be studied in the chiral limit by using the auxiliary-field effective action \cite{Haymaker:1990vm,Roberts:1994dr}, which we'll denote by ${\mathcal A}$.  At $T=0$, realistic models of QCD's gap equation support a DCSB (Nambu-mode) solution and a chirally-symmetric (Wigner-mode) solution.\footnote{
The full pattern of solutions to the gap equation is described in Refs.\,\cite{Zong:2004nm,Chang:2006bm,Williams:2006vva,Fischer:2008sp,Wang:2012me}.  However, as explained in Ref.\,\cite{Wang:2012me}, when discussing phase stability it is sufficient to consider only the simplest Nambu and Wigner solutions.}
The difference
\begin{equation}
\label{bagconstant}
{\mathcal B}_{WN}(T):={\mathcal A}[\mbox{\rm Wigner}]-{\mathcal A}[\mbox{\rm Nambu}]
\end{equation}
measures the relative stability of these different modes \cite{Cahill:1985mh}: when ${\mathcal B}_{WN}$ is positive, the Nambu mode is dynamically favoured.  The difference in Eq.\,\eqref{bagconstant} evolves with $T$; and the temperature at which it vanishes defines the critical value, $T=T_c$, for chiral symmetry restoration.
Chiral symmetry restoration and deconfinement are simultaneous in extant DSE studies; viz., $T_d=T_c$ (see, e.g., Refs.\,\cite{Bender:1996bm,Bender:1997jf,Bashir:2008fk,Bashir:2009fv,Mueller:2010ah,Qin:2010pc}).  Hence, following an idea in Ref.\,\cite{Mo:2010zza}, we define
\begin{equation}
\label{tauirT}
\tau_{\rm ir}(T) = \tau_{\rm ir}\, \frac{{\mathcal B}^{1/4}_{WN}(0)}{{\mathcal B}^{1/4}_{WN}(T)}\,,
\end{equation}
which ensures that the confinement length-scale diverges when chiral symmetry is restored in the chiral limit.

Our precise implementation of Eq.\,\eqref{tauirT} is described in connection with Eq.\,\eqref{tauirTF}.  It uses the fact that, with a dressed-quark propagator of the type in Eq.\,\eqref{SpT}, the explicit form of ${\cal B}_{WN}(T)$ is readily evaluated \cite{Roberts:2000aa}:
\begin{eqnarray}
\mathcal{B}_{WN}(T) & = & 2 N_{c} N_{f} \int_{l,dp}
\left\{ \ln{\Big[\frac{p_l^{2} +
M_{N}^{2}} {p_l^{2} + M_{W}^{2}} \Big]} \nonumber  \right. \\
& & \left. + \Big[ \frac{p_l^{2} + m M_{N}}{p_l^{2} +
M_{N}^{2}} -\frac{p_l^{2} + m M_{W}}{p_l^{2} + M_{W}^{2}} \Big]
\right\}  \,,
\label{eq:B}
\end{eqnarray}
where $p_l^2 = \vec{p}^{\,2} + \omega_l^2$ and $N_f=2$, $N_c=3$.  Equation~\eqref{eq:B} possesses an ultraviolet divergence; and our regularisation procedure is explained in connection with Eq.\,\eqref{eq:Breg}.

\begin{figure}[t]
\includegraphics[width=0.9\linewidth]{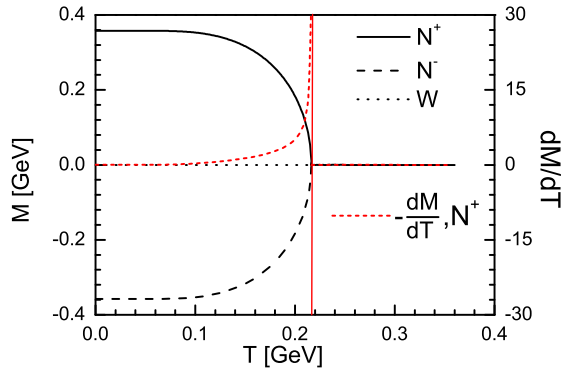}
\caption{\label{fig:Mass-T}
Temperature dependence of the dressed-quark mass in the chiral limit. \emph{Solid curve} -- the standard (positive) Nambu solution, $M_+$; \emph{long dashed curve} -- negative Nambu solution, which necessarily partners the positive solution; \emph{dotted curve} -- Wigner-mode solution; and \emph{short dashed curve} -- $(-dM_+/dT)$.}
\end{figure}

\section{Results}
\label{numerical}
\subsection{Phase transition}
In order to explore chiral symmetry restoration in our symmetry-preserving regularisation of the contact interaction, we solved Eq.\,\eqref{gapactual} in the chiral limit using the values of $\alpha_{\rm IR}$, $\tau_{\rm ir}$, $\tau_{\rm uv}$ specified in Table~\ref{Tab:DressedQuarks}.  The solutions are depicted in Fig.\,\ref{fig:Mass-T}.  It is plain that chiral symmetry is restored via a second-order transition at
\begin{equation}
\label{eq:Tc}
T_c^0 = 0.215 \, {\rm GeV},
\end{equation}
which is the temperature at which the chiral susceptibility, $\chi := -d M/dT$, diverges in the Nambu phase.

Denoting the positive and negative Nambu solutions by $N_\pm$, respectively, then it is not surprising that in the chiral limit
\begin{equation}
\forall T\,: \; \mathcal{B}_{WN_+}(T) - \mathcal{B}_{WN_-}(T) \equiv 0 \,.
\end{equation}

The value of $T_c^0$ in Eq.\,\eqref{eq:Tc} is between 20\% and 40\% too large when compared directly with that obtained in numerical simulations of lattice-regularised QCD \cite{Aoki:2009sc,Gupta:2011wh,Bazavov:2011nk}.  It is notable, however, that $T_c^0 = 0.234 \, m_\rho^0$, where $m_\rho^0=0.919\,$GeV is the model's zero-temperature chiral-limit value for the $\rho$-meson mass.  Measured in these units, our value of $T_c^0$ is between 0\% and 15\% too large, a window that is typical of the rainbow-ladder truncation \cite{Maris:2003vk,Chang:2012cc,Chen:2012tx}.

\begin{figure}[t]
\includegraphics[width=0.9\linewidth]{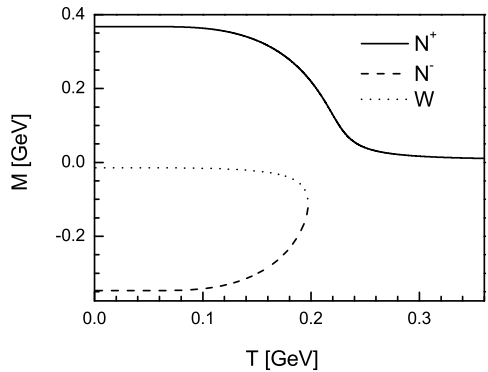}
\caption{\label{fig:MbeyondCL-T}
Temperature dependence of the dressed-quark mass at $m=7\,$MeV, the physical light-quark current-mass in our model.  \emph{Solid curve} -- the standard (positive) Nambu solution, $M_+$; \emph{long dashed curve} -- negative Nambu solution; and \emph{dotted curve} -- Wigner-mode solution.}
\end{figure}

In Fig.\,\ref{fig:MbeyondCL-T} we depict the $T$-dependence of the simplest solutions of the gap equation at the physical value of the light-quark current-mass.  By contrasting Figs.\,\ref{fig:Mass-T} and \ref{fig:MbeyondCL-T}, it becomes evident that the chiral symmetry restoring transition is replaced by a crossover at nonzero current-mass.  In order to implement Eq.\,\eqref{tauirT}, we must ask how then to assign a unique critical temperature for light-quarks with $m\neq 0$?

An answer is suggested by the behaviour of the $N_-$ and $W$ solutions in the Figure; and explained via a thorough consideration of the nature of the gap equation's solutions, as described in Ref.\,\cite{Wang:2012me}.  Expressed simply, in the neighbourhood of $m=0$, DCSB is manifested in the simultaneous existence of three solutions to the gap equation.  When $T$ reaches a value such that just one solution remains, explicit chiral symmetry breaking has come to dominate in the solution of the gap equation.  For $m\gtrsim 0$, therefore, $T_c^m$ is defined as the merging temperature of the $N_-$ and $W$ solutions, which may readily be located.  It is the common temperature at which the chiral susceptibility diverges when evaluated with the $W$ and $N_-$ phases or, equivalently, the solution set
\begin{equation}
\label{defTcm}
\{T_c^m\} = \{ T > 0 | \mathcal{B}_{WN_-}(T) = 0\} \,.
\end{equation}

\begin{figure}[t]
\includegraphics[width=0.95\linewidth]{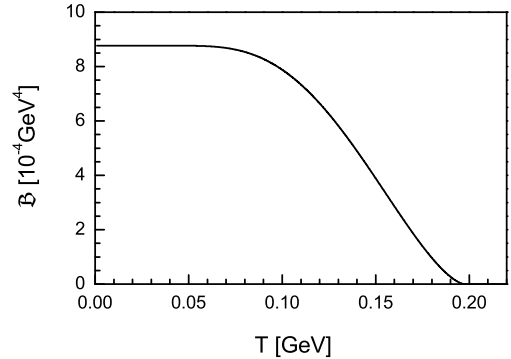}
\includegraphics[width=0.90\linewidth]{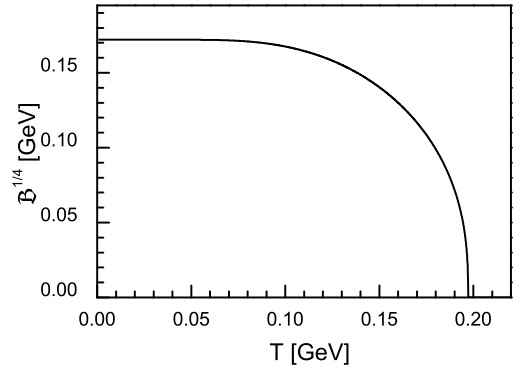}\hspace*{2em}
\caption{\label{fig:BR}
\emph{Upper panel} -- Temperature dependence of the energy-density difference $\mathcal{B}_{WN_-}(T)$ at $m=7\,$MeV.  \emph{Lower panel} -- Fourth-root of that quantity: as noted elsewhere \cite{Cahill:1985mh}, $\mathcal{B}^{1/4}_{WN_-}(T=0)=0.17\,$GeV is commensurate with the energy-difference assumed in bag-like models of baryons.}
\end{figure}

In order to define this set, the integral must be regularised.  We accomplish this by first noting that
\begin{equation}
\frac{\delta}{\delta m} \mathcal{B}_{WN_-}(T) = \langle \bar q q \rangle_{W} - \langle \bar q q \rangle_{N_-} \,,
\end{equation}
where
\begin{equation}
\langle \bar q q \rangle_{P} = -4 N_c N_f \int_{l dp}\frac{M_P}{p_l^2 + M_P^2}\,,
\end{equation}
an expression which can be regularised via Eq.\,\eqref{eq:regulator}.  Now, at each temperature there is a current-quark mass, $m_c$, such that the $W$ and $N_-$ solutions merge and hence $\langle \bar q q \rangle_{W}^{m_c}= \langle \bar q q \rangle_{N_-}^{m_c}$.  Thus the regularised energy-density difference at current-mass $m$ may be reconstructed as
\begin{equation}
\label{eq:Breg}
\mathcal{B}_{WN}(T) = \int_{m_c}^{m}\! dt \left[
\langle \bar q q \rangle_{W}^{m=t} - \langle \bar q q \rangle_{N_-}^{m=t}
\right]\,.
\end{equation}

At $m=7\,$MeV, as illustrated in Fig.\,\ref{fig:BR}, Eq.\,\eqref{defTcm} yields
\begin{equation}
\label{deconfinementT}
T_c^m = 0.197 \,{\rm GeV} = 0.212 \, m_\rho < T_c^0 \,.
\end{equation}
The behaviour illustrated in the Figure does not depend sensitively on the interaction chosen: it is typical of a second-order symmetry-restoring transition (compare, e.g., the results in Refs.\,\cite{Blaschke:1997bj,Bender:1997jf,Zong:2002rh,Chang:2005ay,Yang:2007zzp,Mo:2010zza}).

We note that, evaluated with the $N_+$ solution, the chiral susceptibility is maximal at
\begin{equation}
T_\chi^m = 0.221\,{\rm GeV}=0.238 \, m_\rho\, = 1.12 \,T_c^m.
\end{equation}

The preceding discussion clarifies Eq.\,\eqref{tauirT} and enables a concrete implementation; viz., for light quarks:
\begin{equation}
\label{tauirTF}
\tau_{\rm ir}^m(T) = \tau_{\rm ir}\, \frac{{\cal B}^{1/4}_{WN_-}(0)}{{\cal B}^{1/4}_{WN_-}(T)}\,.
\end{equation}
All results described below are obtained by using this $T$-dependent infrared length-scale in the computation of the functions $\mathcal C^{\rm ir}$, $\mathcal C_1^{\rm ir}$, $\mathcal C_2^{\rm ir}$.

\begin{figure}[t]
\includegraphics[width=0.9\linewidth]{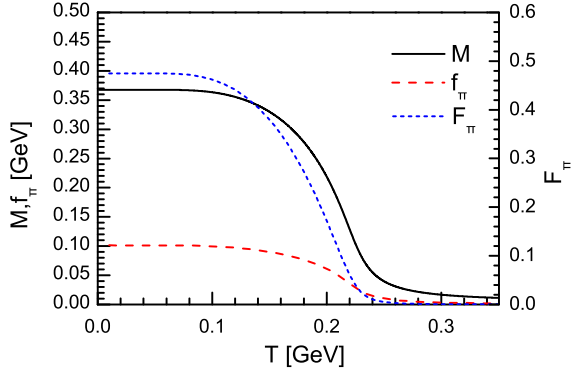}
\caption{\label{fig:dscb}
Calculated $T$-dependence of several quantities that are commonly used to illustrate the evolution of hadron properties through the chiral symmetry restoring transition: \emph{Solid curve} -- dressed-quark mass;
\emph{dotted curve} -- pseudovector component of the pion's Bethe-Salpeter amplitude, $F_{\pi}$, Eq.\,\protect\eqref{piBSA};
and \emph{dashed curve} -- pion's leptonic decay constant $f_\pi$, Eq.\,\protect\eqref{fpiT}.}
\end{figure}

In Fig.\,\ref{fig:dscb} we depict the temperature dependence of quantities that may all be considered as equivalent chiral order parameters.  The novelty, perhaps, is $F_\pi$, which is the pseudovector component of the pion's Bethe-Salpeter amplitude, Eq.\,\eqref{piBSA}.  As remarked above: a pseudoscalar meson must possess components in its Bethe-Salpeter amplitude that may be described as pseudovector in character \cite{Maris:1997hd}; and these pieces materially influence the $T=0$ physics of pseudoscalar mesons \cite{Maris:1998hc,GutierrezGuerrero:2010md,Roberts:2010rn,Chang:2012cc,Chen:2012tx}.  With increasing temperature, however, the strength of these components diminishes until, at $T_c^0$ in the chiral limit, they disappear \cite{Maris:2000ig}.  It is that outcome which forces $f_\pi$ to vanish.  This is abundantly clear when analysing the contact interaction: consider Eqs.\,\eqref{bsefinal0}, \eqref{pionkernel} and recognise that the  expression for $f_\pi$, Eq.\,\eqref{fpiT}, is proportional to the equation for $F_\pi$.  That $F_\pi$ is equivalent to $M$ as an order parameter is also plain: ${\mathcal K}_{EF} \propto m_\pi^2 \propto M$ [see Eq.\,\eqref{KpiEF}] and hence the driving term for $F_\pi$ vanishes with $M$.

\begin{figure}[t]
\includegraphics[width=0.9\linewidth]{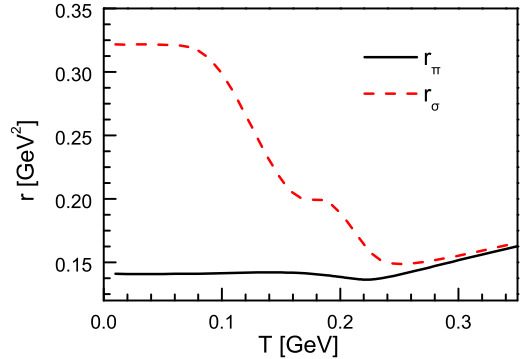}
\caption{\label{fig:rho0}
$T$-dependence of the residues of the $\pi$- and $\sigma$-mesons in, respectively, the pseudoscalar and scalar vertices: \emph{solid curve} -- $\mathpzc r_\pi$ in Eq.\,\protect\eqref{residuepi}; and \emph{dashed curve} -- $\mathpzc r_\sigma$ in Eq.\,\protect\eqref{residuesigma}.}
\end{figure}

In Fig.\,\ref{fig:rho0} we report the temperature dependence of the residues connected with the $\pi$- and $\sigma$-meson poles in, respectively, the inhomogeneous pseudoscalar and scalar vertices.  Qualitatively, the behaviour is similar to that depicted in Fig.\,6 of Ref.\,\cite{Maris:2000ig}: chiral symmetry is restored.  Herein, however, the $T=0$ magnitude of the splitting $\mathpzc r_\sigma-\mathpzc r_\pi$ is greater because we have modified the $\sigma$-meson BSE by the inclusion of $g_{\rm SO}$ in Eq.\,\eqref{gSOT}.  The behaviour of $\mathpzc r_\sigma$ in the neighbourhood of $T_c^m$ is an artefact arising from interference between $g_{\rm SO}(T)$ and $\tau_{\rm ir}(T)$.

\subsection{Screening masses: mesons and diquarks}
We are now in a position to compute and report the $T$-dependent screening masses for the two-valence-body systems whose Bethe-Salpeter equations are detailed in Secs.\,\ref{sec:mesons}, \ref{sec:qqBSA}.

\subsubsection{Screening in $J=0$ channels}
\begin{figure}[t]
\includegraphics[width=0.9\linewidth]{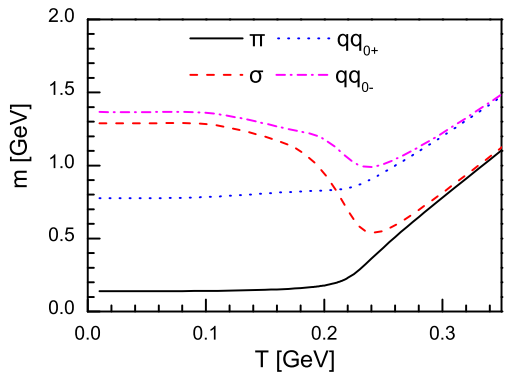}
\caption{\label{fig:debye0}
Screening masses of mesons and diquarks with $J=0$:
\emph{solid curve} -- pion; \emph{dashed curve} -- $\sigma$-meson described in Sec.\,\protect\ref{scalarpseudovector}, the pion's parity partner for $N_f=2$; \emph{dotted curve} -- $J^P=0^+$ diquark correlation; and dot-dashed curve --
$J^{P}=0^{-}$ diquark.}
\end{figure}

In Fig.\,\ref{fig:debye0} we plot the $T$-evolution of the screening masses associated with $J=0$ systems.  At $T=0$, our implementation of the contact interaction produces the following bound-state inertial masses \cite{Chen:2012qr}:
\begin{equation}
\begin{array}{lcccc}
                 & \pi & \sigma & [qq]_{0^+} & [qq]_{0^-}\\
 \mbox{mass\;(GeV)} & 0.14 & 1.29 & 0.78 & 1.37
\end{array}\,.
\end{equation}
Plainly, DCSB is expressed strongly in this part of the spectrum through a large splitting between parity partners.  The splitting persists in the screening masses.  They are approximately insensitive to temperature until $T=T_c^m=T_d$, beyond which value the screening masses of the parity-partner correlations evolve rapidly toward near degeneracy: apart from current-mass effects, they may be called equal for $T\gtrsim 1.3 \, T_c^m$.

Since our confinement mass-scale vanishes for $T>T_c^m=T_d$; i.e., $\Lambda_{\rm ir}(T>T_c^m)=0$, it is noteworthy that the correlations persist on this domain.  Thus, deconfinement is not expressed in the absence of strong correlations in these channels.  This is because fermions at nonzero temperature are characterised by an additional mass-scale: $\omega_0 = \pi T$; and so long as $2[\omega_0 + M(T)]$ exceeds the correlation's screening mass, the correlation will persist.  These features were exposed in Ref.\,\cite{Maris:2000ig}.

Considering the construction described in connection with Eq.\,\eqref{BeginWightmann}, one should expect a dramatic expression in the real-time propagator of the discontinuous derivative exhibited at $T_c^0$ by the lightest chiral-limit screening masses in the $\pi$ and $\sigma$ channels (see Fig.\,3 in Ref.\,\cite{Maris:2000ig}), and its smoother remnant at $m>0$, which is evident in Fig.\,\ref{fig:debye0}.

\subsubsection{Bound states?}
\label{sec:bs}
Given these observations, we judge that the behaviour of the screening masses does not necessarily mean that bound-states continue to exist in any given channel at $T>T_c$.

There are illustrative examples; e.g., Ref.\,\cite{Chang:2008ec}, which considers the scalar and pseudoscalar channels.  In ideal rainbow-ladder truncation (i.e., with $g_{\rm SO}=1$), the inertial mass of the scalar-meson is $m_\sigma = 2M$ in the chiral limit, and the pseudoscalar mass is zero.  As the $T=0$ interaction strength is reduced to some critical value (e.g., $\alpha_{\rm IR}$ is reduced), the dressed-quark mass $M\to 0$ and hence $m_\sigma \to 0$, thus becoming degenerate with $m_\pi = 0$.  As the interaction strength is reduced still further, the bound-states disappear.  It is possible, therefore, that when some mechanism suppresses the interaction-strength to a sufficiently large extent, dynamical mass generation is impossible and no true bound-states are supported.

With these observations in mind, we conjecture that, when they may reasonably be defined, the inertial masses of all hadron bound-states are proportional to $M(T)$ for $T<T_d$; and that no bound-states persist above $T_d$.  This suggestion is less surprising if one states it thus: as the critical temperature characterising a second-order phase boundary is approached from below, all correlation lengths diverge and real-time correlation functions acquire power-law behaviour.  In our case, the primary correlation length is $\xi(T) \simeq \tau_{\rm ir}(T) \simeq 1/M(T)$, the divergence of which forces all related correlation-lengths -- the inverse of hadron inertial-masses -- to diverge as well.  Notably, for states with a significant hadronic width, a vanishing mass might be very difficult to distinguish empirically from marked spectral broadening: $M \to m - i \omega$, with  $\omega/m \gg 1$.  We note in connection with the latter that since rainbow-ladder truncation omits resonant contributions to bound-state kernels, it is likely to become a quantitatively inaccurate approximation if used to compute the dynamical evolution of bound-states in a medium with copious numbers of thermal pions.

The possibility that hadron inertial-masses decrease in-medium has long excited interest \cite{Brown:1991kk}; and many analyses have explored this possibility \cite{David:2006sr,Hayano:2008vn,Leupold:2009kz}.  At present the notion is empirically neither confirmed nor invalidated, although an explanation of data does not require this effect in its simpler forms \cite{Rapp:2012zq}.  A reduction in hadron inertial masses in the neighbourhood of $T_c$ is consistent with results from some applications of sum rules but spectral broadening via hadron-hadron interactions in-medium is also very important \cite{Zschocke:2002mn,Kwon:2010fw,Hilger:2010cn}.  Much remains to be learnt in this area, at least in connection with the role and manifestation of gluon-quark dynamics.

\begin{figure}[t]
\includegraphics[width=0.9\linewidth]{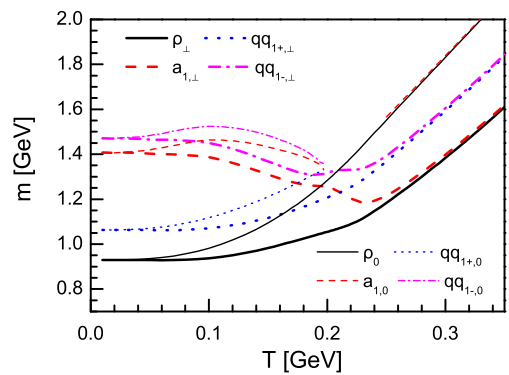}
\caption{\label{fig:debye1}
Screening masses of mesons and diquarks with $J=1$:
\emph{solid curve} -- $\rho$-meson; \emph{dashed curve} -- $a_1$-meson; \emph{dotted curve} -- $J^P=1^+$ diquark correlation; and dot-dashed curve --
$J^{P}=1^{-}$ diquark.  Transverse modes are traced with thick lines and longitudinal models with thin lines.}
\end{figure}

\subsubsection{Screening in $J=1$ channels}
We plot the $T$-evolution of screening masses associated with $J=1$ systems in Fig.\,\ref{fig:debye1}.  At $T=0$, our implementation of the contact interaction produces \cite{Chen:2012qr}:
\begin{equation}
\begin{array}{lcccc}
                 & \rho & a_1 & \{qq\}_{1^+} & \{qq\}_{1^-}\\
 \mbox{mass\;(GeV)} & 0.93 & 1.38 & 1.06 & 1.45
\end{array}\,.
\end{equation}
For $T>0.3\,T_c^m$ a separation is apparent between the screening masses of the transverse and longitudinal modes.  The subsequent behaviour of the screening masses for the transverse modes follows the pattern set by $J=0$ systems.  They are weakly sensitive to temperature within the confinement domain, a result found previously in the algebraic model of Ref.\,\cite{Maris:1997eg} (see Eq.\,(21) therein); and chiral symmetry restoration is again evident for $T\gtrsim 1.3 \, T_c^m$, apart from current-mass effects.

The longitudinal modes behave differently, however.  Their screening masses all increase markedly with temperature, a result readily understandable from Eq.(22) in Ref.\,\cite{Maris:1997eg}, such that, with the exception of $m_{\rho^\parallel}$, they are greater than $2\omega_0$ at the deconfinement temperature, $T_d=T_c^m$ in Eq.\,\eqref{deconfinementT}.  In these three channels, owing to the additional repulsion produced by the $\mathcal R^{\rm iu}(T)$ term in the Bethe-Salpeter kernels [see, e.g.,  Eqs.\,\eqref{a1parallel}, \eqref{Ka1parallel}], this is sufficient to dissolve the correlations.

\begin{figure}
\includegraphics[width=0.9\linewidth]{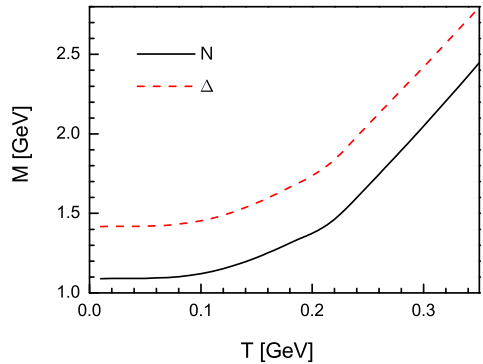}
\caption{\label{fig:mB}
Screening masses of the nucleon and $\Delta$ dressed-quark cores: \emph{solid curve} -- nucleon; and \emph{dashed curve} -- $\Delta$.}
\end{figure}

\subsection{Screening masses: nucleon and $\Delta$}
Having determined the behaviour of the screening masses and Bethe-Salpeter amplitudes of the diquark correlations, we are now positioned to compute and report the temperature dependence of the dressed-quark cores of the nucleon and $\Delta$ as they are described by the Faddeev equations detailed in Sec.\,\ref{subsec:baryon}.  The $T=0$ inertial masses are listed and discussed, respectively, in connection with Eqs.\,\eqref{MassN}, \eqref{Deltamass}.  Their evolution into temperature-dependent screening masses is depicted in Fig.\,\ref{fig:mB}: correlations persist in both channels for all values of $T$; and for $T\gtrsim T_\chi^m$, the screening masses increase linearly with temperature but always lie below $3\pi T$.

\begin{figure}
\includegraphics[width=0.9\linewidth]{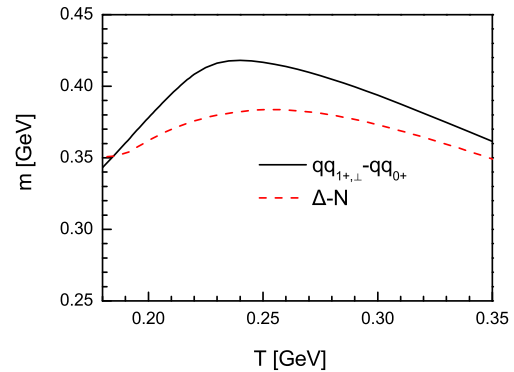}
\caption{\label{fig:massdifference}
Difference between screening masses: \emph{solid curve} -- $\Delta$ and nucleon dressed-quark cores; and \emph{dashed curve} -- axial-vector and vector diquark correlations.}
\end{figure}

It is interesting to analyse the splitting between the $\Delta$ and nucleon screening masses.  In the $T=0$ study of Ref.\,\cite{Roberts:2011cf} it was shown that, as a function of current-mass, $m_\Delta - m_N$ is linearly proportional to the splitting $m_{\{qq\}_{1^+}}-m_{[qq]_{0^+}}$.  Figure~\ref{fig:massdifference} demonstrates that a similar correspondence holds at fixed current-mass with increasing temperature: given the $T$-dependence of $m_{\{qq\}_{1^+}}-m_{[qq]_{0^+}}$, then that of $m_\Delta - m_N$ is approximately the same.

The existence of diquark correlations within baryons is a dynamical outcome of the strong interaction between quarks.  Whether one exploits this feature in order to develop an approximation to the quark-quark scattering matrix, as we do herein, or chooses instead to eschew the simplification it offers, the outcome is the same \cite{Eichmann:2009qa}.  It follows that axial-vector diquark correlations are dominant within the $\Delta$.  The nucleon, on the other hand, possesses both scalar and axial-vector diquarks; and the nucleon's Faddeev amplitude expresses the relative strength of these different correlations within the nucleon.  The $T=0$ result is presented in Eq.\,\eqref{FaddeevAN}: the scalar diquark is found with 72\% probability.  (The significance of this result for the hadron spectrum is described in Ref.\,\cite{Chen:2012qr}.)

It is natural to consider how the ratio evolves with temperature in the correlations that persist above $T_d=T_c^m$.  This is depicted in Fig.\,\ref{fig:NucleonAmplitude}: whilst there is a perceptible evolution in the apportionment of strength between the two axial-vector diquark components, with the relative probability switching at $T=T_d$, the probability of finding a scalar diquark correlation is almost temperature independent.  This is a curious outcome, which owes to a relative similarity in the evolution of both $E_{0^+}$, $E_{1^+}$ and $m_{0^+}$, $m_{q^+}$, as illustrated by Fig.\,\ref{fig:massdifference}.  In contrast, Fig.\,\ref{fig:dscb} shows that the structures defining the pseudoscalar correlation, and hence the scalar diquark, both change significantly above $T_d$.  It will be interesting to learn whether the behaviour in Fig.\,\ref{fig:NucleonAmplitude} survives a more sophisticated treatment of the $T\neq 0$ Faddeev equation.

One might also ask about parity partners in the baryon sector.  Based upon the expressions in Ref.\,\cite{Chen:2012qr} and their relationship to those derived herein, we anticipate that chiral symmetry restoration will lead to degeneracy between $J^P = \frac{1}{2}^+, \frac{1}{2}^-$ screening masses and, separately, $J^P = \frac{3}{2}^+, \frac{3}{2}^-$ masses.

\section{Goldberger-Treiman Relations}
\label{sec:GTR}
In-vacuum, the axial-vector vertex, $\Gamma_{5\mu}(k_+,k)$, $k_+= k+Q$, is the solution of
\begin{eqnarray}
\nonumber
\lefteqn{\Gamma_{5\mu}(k_+,k) = \gamma_5\gamma_\mu} \\
&&  - \frac{16\pi\alpha_{\rm IR}}{3 m_G^2}\int\! \frac{d^4t}{(2\pi)^4}
\gamma_\alpha S(t+Q) \Gamma_{5\mu}(Q) S(t) \gamma_\alpha\,. \label{IBSEAV}
\end{eqnarray}
It satisfies the axial-vector Ward-Green-Takahashi identity, which reads, in the chiral limit:
\begin{equation}
\label{avwti}
P_\mu \Gamma_{5\mu}(k_+,k) = S^{-1}(k_+) i \gamma_5 + i \gamma_5 S^{-1}(k)\,.
\end{equation}

\begin{figure}
\includegraphics[width=0.9\linewidth]{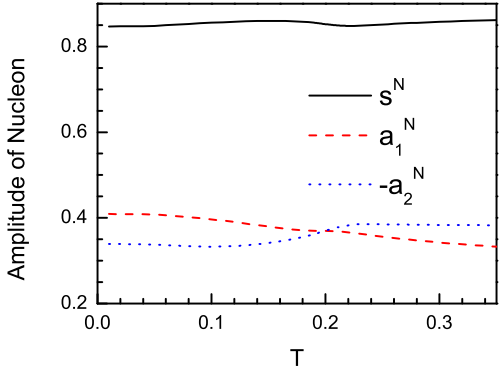}
\caption{\label{fig:NucleonAmplitude}
Evolution of the nucleon's Faddeev amplitude with temperature: \emph{solid curve} -- scalar diquark component; and \emph{dashed, dotted curves} -- the two distinct axial-vector diquark structures.  [See Eqs.\,\protect\eqref{psiN}, \protect\eqref{psiNdetail}.]}
\end{figure}

Translating the general relations in Ref.\,\cite{Maris:1997hd} to our symmetry-preserving formulation of the contact interaction, Eq.\,\eqref{avwti} entails the following chiral-limit quark-level Goldberger-Treiman relations:
\begin{eqnarray}
f_\pi^0 E^0_\pi &=& M^0\,,\\
M^0 g_{Aq}^0 + 2 f_\pi^0  F_\pi^0 &=& M^0 = f_\pi^0 E_\pi^0\,,
\end{eqnarray}
where, as above, the superscript ``0'' denotes a quantity computed in the chiral limit and we have used the fact that, in this limit and in the neighbourhood of $Q^2=0$, the axial-vector vertex has the general form:
\begin{eqnarray}
\Gamma_{5\mu}^0(k_+,k) & = & \gamma_5 \gamma_\mu F_R^0 (Q) + \frac{Q_\mu}{Q^2} 2 f_\pi^0 \Gamma_\pi^0(Q)\,.
\end{eqnarray}
Here, $\Gamma_\pi^0(Q)$ is the canonically normalised pion Bethe-Salpeter amplitude and we have defined a dressed-quark axial-charge \cite{Chang:2012cc}
\begin{equation}
\label{qaxialcharge}
g_{Aq}^0 = F_R^0 (Q=0)\,.
\end{equation}

With the loss of $O(4)$ invariance at nonzero temperature, the contact-interaction axial-vector vertex takes the general form ($Q_0 = \{\vec{Q},0\}$)
\begin{equation}
\label{G5muT}
\Gamma_{5\mu}(Q_0) =
\gamma_5
\left\{
\begin{array}{l}
\gamma_4 F^\parallel(Q_0), \\
\vec{\gamma} F_R^\perp(Q_0) + 2 \vec{\gamma}_\parallel  F_I^\perp(Q_0) + 2 i \frac{\vec{Q}}{\vec{Q}^2} E^\perp(Q_0),
\end{array}
\right.
\end{equation}
where 
$\vec{\gamma} =: \vec{\gamma}_\perp + \vec{\gamma}_\parallel$ with $\vec{Q}\cdot \vec{\gamma}_\parallel =\vec{Q}\cdot \vec{\gamma}$.  In this case, the axial-vector Ward-Green-Takahashi identity does not place a tight constraint on $F^\parallel(Q_0)$.  However, the existence at $T\neq 0$ of a pseudoscalar correlation with zero screening mass entails:
\begin{subequations}
{\allowdisplaybreaks
\label{GTrqT}
\begin{eqnarray}
E^{\perp 0}(0) &=& f_\pi^0 E_\pi^0(0) = M^0 \,,\\
M^0 F_I^{\perp 0}(0) & = & f_\pi^0 F_\pi^0(0)\,,\\
M^0 F_R^{\perp 0}(0) + 2 f_\pi^0 F_\pi^0(0) &=& M^0 = f_\pi^0 E_\pi^0(0)\,;
\label{gAqperpGTr}
\end{eqnarray}}
\end{subequations}
\hspace*{-0.4\parindent}and vice-versa.

\begin{figure}
\includegraphics[width=0.9\linewidth]{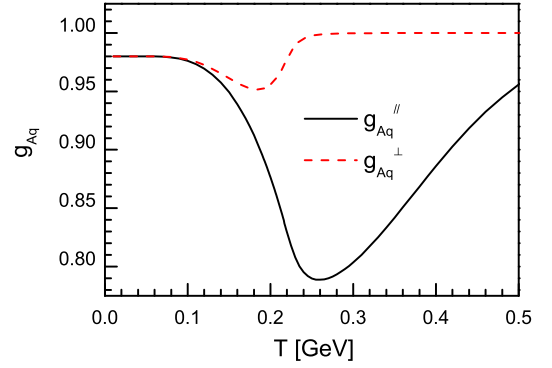}
\caption{\label{fig:gAq}
Temperature evolution of the dressed-quark axial-charges in Eqs.\,\protect\eqref{eq:gAqT}, evaluated with $m=7\,$MeV: \emph{solid curve} -- $g_{Aq}^{\parallel}$; and \emph{dashed curve} -- $g_{Aq}^{\perp}$.}
\end{figure}

As indicated in Eq.\,\eqref{qaxialcharge}, the regular parts of the axial-vector vertex may be identified as axial-charges of a dressed-quark.  The $T$-dependence of these charges is described by the following formulae:
\begin{subequations}
\label{eq:gAqT}
\begin{eqnarray}
g_{Aq}^{\perp}(T) &=&  F^\parallel(0) = \frac{1}{1 + \mathpzc K_{a_1^\perp}(0) },\\
g_{Aq}^{\parallel}(T) &=& F_R^\perp(0) = \frac{1}{1 + \mathpzc K_{a_1^\parallel}(0) },
\end{eqnarray}
\end{subequations}
with the kernels given in Eqs.\,\eqref{Ka1perp}, \eqref{Ka1parallel}; and displayed in Fig.\,\ref{fig:gAq}.  The behaviour is consistent with the $T\neq 0$ Goldberger-Treiman relations, discussed in association with Eqs.\,\eqref{G5muT}, \eqref{GTrqT}.  The charges are identical and less-than one for $\pi T \lesssim M(0)$; and also essentially independent of $T$ on this domain (consistent with Ref.\,\cite{Eletsky:1993hp}, although the context therein is different).  However, $g_A^\perp$ and $g_A^\parallel$ become distinct on $\pi T > M(0)$.  The transverse charge remains below one until $T \approx T_\chi^m$ whereafter it rapidly approaches unity, as dictated by Eq.\,\eqref{gAqperpGTr} and the restoration of chiral symmetry.  Like the screening masses, however, the longitudinal axial-charge is far more sensitive to temperature than $g_{Aq}^\perp$.  At $T=T_\chi^m$, $g_{Aq}^\parallel$ has dropped 20\% from its $T=0$ value.  With increasing temperature thereafter, it approaches unity from below; but only slowly: e.g., even at $T=2T_\chi^m$ it has not returned to its $T=0$ value.

The connection in-vacuum between the dressed-quark's axial-charge and that of the nucleon is discussed in Ref.\,\cite{Chang:2012cc}.  We revisit that here in the context of nonzero temperature.  In this case the free-field nucleon spinor is defined via
\begin{subequations}
\label{eq:DiracNucleon}
\begin{eqnarray}
0 &= &[i\vec{\gamma}\cdot \vec{P} + i \gamma_4 \omega_n + M_N ]\, u_n(\vec{P}),\\
&=& \bar u_n(\vec{P})\, [i\vec{\gamma}\cdot \vec{P} + i \gamma_4 \omega_n + M_N ] ,
\end{eqnarray}
\end{subequations}
where $M_N$ is the fermion's inertial mass.  It follows that
\begin{eqnarray}
\nonumber
&& \bar u_n(\vec{P}^\prime)\, \gamma_5 \vec{\gamma}\cdot (\vec{P^\prime}-\vec{P}) \, u_m(\vec{P})\\
&=& \bar u_n(\vec{P}^\prime)\gamma_5 [ -\gamma_4 (\omega_n - \omega_m) - 2 i M_N ] u_m(\vec{P})\,.
\end{eqnarray}

Now consider the $T\neq 0$ extension of the nucleon's axial-vector current with momentum $Q_0$ entering the vertex:
\begin{eqnarray}
\vec{J}_5(\vec{Q}) &= & \bar u_n(\vec{P}^\prime)\, \gamma_5 [
   \vec{\gamma} \, g_A^\perp(\vec{Q}^2)
+   \vec{Q}\, g_P(\vec{Q^2}) ] u_n(\vec{P}).
\end{eqnarray}
If one specialises to the chiral limit, then, analogous to the dressed-quark case, the existence of pseudoscalar correlation with zero screening mass entails
\begin{eqnarray}
\nonumber && \bar u_n(\vec{P}^\prime)\, \gamma_5   g_P^0(\vec{Q^2})  u_n(\vec{P})\\
&\stackrel{\vec{Q}^2\sim 0}{=} & 2 \frac{f_\pi^0 }{\vec{Q}^2}
g_{\pi NN}^0(0) \, \bar u_n(\vec{P}^\prime)\, i\gamma_5  u_n(\vec{P})\,,
\end{eqnarray}
where $g_{\pi NN}^0(0)$ is a normalisation factor that expresses the $\vec{Q}^2=0$ value of the seven-point function $\bar u_n(\vec{P}^\prime)\, i\gamma_5  u_n(\vec{P})$.  As a consequence of the nucleon-level axial-vector Ward-Green-Takahashi identity, one has
\begin{eqnarray}
0 & = &  i \vec{Q}\cdot \vec{J}_5^0(\vec{Q}) \\
&=&
\bar u_n(\vec{P}^\prime)\, i\gamma_5 [
   \vec{\gamma} \cdot\vec{Q} \, g_A^{\perp 0}(\vec{Q}^2)
+  \vec{Q}^2\, g_P^0(\vec{Q^2}) ] u_n(\vec{P}) \rule{1.5em}{0ex} \\
&\stackrel{\vec{Q}^2\sim 0}{=} & 2 \bar u_n(\vec{P}^\prime)\, \gamma_5
[  M_N g_A^{\perp 0}(0) -  f_\pi^0 g_{\pi NN}^0 ]u_n(\vec{P})\,;
\end{eqnarray}
and hence, in the chiral limit, at all values of temperature:
\begin{equation}
M_N^0 g_A^{\perp 0}(0) =  f_\pi^0 g_{\pi NN}^0(0)\,.
\end{equation}

In the chiral limit, $f_\pi^0 = 0 $ at $T_c^0$.  As with $\Gamma_\pi^0(Q)$, $\forall \,T$ the normalised value of $\bar u_n(\vec{P}^\prime)\, i\gamma_5  u_n(\vec{P})$ is finite at $\vec{Q^2}=0$; i.e., $0<g_{\pi NN}^0(0)<\infty$.  Consequently,
\begin{equation}
\label{NGTrT}
\lim_{T \to (T_c^0)^-} M_N^0 g_A^{\perp 0}(0) = 0\,.
\end{equation}

Following the reasoning in Ref.\,\cite{Chang:2012cc}, since $g_{Aq}^{\perp 0}$ is always nonzero and, indeed, $g_{Aq}^{\perp 0}(T=T_c^0) \approx 1$, Eq.\,\eqref{NGTrT} is not achieved by changes at the level of the dressed-quark--axial-vector vertex.  Hence a vanishing of $g_A^{\perp 0}(T_c^0)$ would require extraordinary and precise cancellations amongst the terms that constitute the axial-charge matrix element associated with the correlation in the nucleon channel; namely, between the various contributions arising from the angular momentum correlations within the Faddeev amplitude.  Owing to the power of symmetries in quantum field theory, this is conceivable.  However, it is unlikely given the behaviour in Fig.\,\ref{fig:NucleonAmplitude}; i.e., that the amplitude describing the $T\neq 0$ correlation in the nucleon channel is only weakly sensitive to $T$.  It is more probable, therefore, as argued in Ref.\,\cite{Chang:2012cc}, that Eq.\,\eqref{NGTrT} is achieved via dissolution of the nucleon bound-state at a point of coincident chiral symmetry restoration and deconfinement, with
\begin{equation}
\label{NMT}
\lim_{T \to (T_c^0)^-} M_N^0  = 0\,,
\end{equation}
so that, beyond $T_c$, $g_A^{\perp 0}$ represents just the normalisation of a seven-point function that is associated with a strong screening-correlation but not a bound-state.  The discussion in Sec.\,\ref{sec:bs} anticipates Eq.\,\eqref{NMT}.

The above discussion is dubious in the neighbourhood of $T_c^0$ if the inertial mass of the nucleon acquires a large imaginary part within this domain; i.e., if $M_N^0\to m_N^0 - i \omega_N^0$, with $\omega_N^0/m_N^0 \gtrsim 1$.  In this case, Eqs.\,\eqref{eq:DiracNucleon} become poor approximations.  In fact, there is no sense in which one may employ notions of an asymptotic nucleon state and the definition of each of the vertices employed above must be revised significantly.  But this is also the content of Ref.\,\cite{Chang:2012cc}: under these conditions, the Goldberger-Treiman relation is made moot by bound-state dissolution.

\section{Summary and Perspective}
\label{summary}
Working at leading-order in a symmetry-preserving truncation scheme for the Dyson-Schwinger equations, we extended a confining formulation of a vector$\times$vector contact-interaction to nonzero temperature.  This framework proved useful in the study of a wide range of phenomena at $T=0$, including hadron masses and form factors. 
We therefore expect that, interpreted judiciously, results obtained at $T>0$ should represent a fair guide to related hadron properties on this new domain.

In formulating the interaction at nonzero temperature, our treatment of the gap and Bethe-Salpeter equations is standard.  However, our formulation of the baryon Faddeev equations is novel.  Although, in common with previous continuum studies, our treatment fails to express the full complexity of $J=\frac{1}{2},\frac{3}{2}$ states at $T>0$, it does improve upon preceding analyses in a number of ways.  For example, by: including axial-vector diquarks in addition to the scalar correlations; improving materially upon the implementation of the widely-used static approximation for the Faddeev kernels; and allowing dynamically for the expression of chiral symmetry restoration and, importantly, deconfinement.

Turning to the results, via the gap equation we found a second-order chiral-symmetry restoring transition at $T=T_c^0 = 0.23 \,m_\rho$ in the chiral limit, which becomes a cross-over at nonzero current-mass.  Notwithstanding this change at $m\neq 0$, by capitalising on the modern understanding of the gap equation and the nature of its solutions, 
we were still able to define a single temperature, $T_c^m=0.21\,m_\rho$, whereat dynamical chiral symmetry breaking (DCSB) is no longer effective.

We thereafter explained and implemented a dynamical mechanism that ensured deconfinement at $T_d=T_c^m$.  It is distinguished by the feature that, whilst no kernel or process exhibits production thresholds for coloured states when $T < T_d$, all do for larger temperatures.  Despite this, in all $T=0$ bound-state channels, strong correlations persist for $T > T_d$.  Furthermore, in the mass spectrum defined by the associated screening masses, degeneracy between parity partners is apparent for $T \gtrsim 1.3\,T_c^m$.

Since we retained axial-vector diquark correlations, we were able to simultaneously study nucleon and $\Delta$-baryon properties.  The splitting between screening masses in these channels evolves with temperature in a manner that is approximately proportional to the splitting between the axial-vector and scalar diquark masses.   Curiously, we found that the scalar-diquark content of the correlation in the nucleon channel is almost independent of temperature: it is 72\% on $T \lesssim 2 T_c$, which is the highest temperature considered herein.

Notwithstanding our results for screening masses and the associated correlation amplitudes, we argued that there are reasons (amongst them, the nucleon's Goldberger-Treiman relation) to suspect that, when they may reasonably be defined, bound-state inertial masses vanish as $T\to T_d^-$; and that this is a signal of bound-state dissolution at the deconfinement temperature.

In closing we reflect briefly on the question of the veracity of our results.  Our formulation is internally consistent and more systematic than preceding continuum studies of the properties addressed herein.  We have furthermore eliminated two parameters used in earlier formulations of the contact-interaction baryon Faddeev equations at $T=0$ and left the other three untouched.  The material weakness is that the rainbow-ladder truncation omits contributions to the gap and bound-state kernels which might fairly be described as meson-loop corrections.  However, whilst such corrections are necessary in order to reliably determine critical exponents associated with the transitions (see Ref.\,\cite{Roberts:2000aa}, p.\,S49, and Refs.\,\cite{Holl:1998qs,Fischer:2011pk}), we do not expect them to have a material effect on the quantities we've focused upon herein.  This assumption can be checked.  For the present, however, we note that our predictions, e.g., regarding the persistence of correlations above $T_c$ and the nature of chiral symmetry restoration and deconfinement, are broadly consistent with analyses of contemporary lattice simulations in those instances where a sensible comparison is possible.  It is interesting now to turn this analysis toward the questions we've raised herein; e.g., developing a continuum connection between the screening masses and the real-time structure of spectral functions.

\begin{acknowledgments}
We thank C.~Chen, T.~Kl\"ahn, R.~Rapp, D.\,H.~Rischke and A.~Sedrakian for valuable comments and explanations.
This work was supported by:
the National Natural Science Foundation of China under contract Nos.\ 10935001, 11075052 and 11175004;
the National Key Basic Research Program of China under contract No.\ 2013CB834400;
Forschungszentrum J\"ulich GmbH;
and
U.\,S.\ Department of Energy, Office of Nuclear Physics, contract no.~DE-AC02-06CH11357.
\end{acknowledgments}

\appendix
\section{Contact interaction}
\label{sec:contact}
The key elements in our analysis are the dressed-quark propagator, and the meson and diquark Bethe-Salpeter amplitudes.  All are completely determined once the quark-quark interaction kernel is specified.  We use
\begin{equation}
\label{njlgluon}
g^2 D_{\mu \nu}(p-q) = \delta_{\mu \nu} \frac{4 \pi \alpha_{\rm IR}}{m_G^2}\,,
\end{equation}
where $m_G=0.8\,$GeV is a gluon mass-scale typical of the one-loop renormalisation-group-improved interaction detailed in Ref.\,\cite{Qin:2011dd}, and the fitted parameter $\alpha_{\rm IR} = 0.93 \pi$ is commensurate with contemporary estimates of the zero-momentum value of a running-coupling in QCD \cite{Aguilar:2009nf,Oliveira:2010xc,Aguilar:2010gm,Boucaud:2010gr,Pennington:2011xs,Wilson:2012em}.  We embed Eq.\,\eqref{njlgluon} in a rainbow-ladder truncation of the DSEs.  This means
\begin{equation}
\label{RLvertex}
\Gamma_{\nu}(p,q) =\gamma_{\nu}
\end{equation}
in the gap equation and in the subsequent construction of the Bethe-Salpeter kernels.

Whilst the interaction in Eq.\,(\ref{njlgluon}) may be viewed as being inspired by models of the Nambu--Jona-Lasinio type \cite{Nambu:1961tp}, our treatment is atypical.  Used to build a rainbow-ladder truncation of the DSEs, Eqs.\,\eqref{njlgluon}, (\ref{RLvertex}) produce results for low-momentum-transfer observables that are directly comparable with those produced by more sophisticated interactions, as illustrated in
Refs.\,\cite{GutierrezGuerrero:2010md,Roberts:2010rn,Roberts:2011cf,Roberts:2011wy,%
Wilson:2011aa,Chen:2012qr,Chen:2012tx}.

\begin{table}[t]
\caption{\label{Tab:DressedQuarks}
Dressed-quark properties, computed from the gap equation and required as input for the Bethe-Salpeter and Faddeev equations, and computed values for in-hadron condensates \protect\cite{Brodsky:2012ku} -- all at $T=0$.
The results are obtained with $\alpha_{\rm IR} =0.93 \pi$ and (in GeV) $\Lambda_{\rm ir} = 0.24\,$, $\Lambda_{\rm uv}=0.905$.  (These parameters take the values determined in the spectrum calculation of Ref.\,\protect\cite{Roberts:2011cf}, which produces $m_\rho=0.928\,$GeV; we assume isospin symmetry throughout; and all dimensioned quantities are listed in GeV.)}
\begin{center}
\begin{tabular*}
{\hsize}
{
c@{\extracolsep{0ptplus1fil}}
c@{\extracolsep{0ptplus1fil}}
c@{\extracolsep{0ptplus1fil}}
c@{\extracolsep{0ptplus1fil}}
c@{\extracolsep{0ptplus1fil}}
c@{\extracolsep{0ptplus1fil}}
c@{\extracolsep{0ptplus1fil}}
c@{\extracolsep{0ptplus1fil}}
c@{\extracolsep{0ptplus1fil}}
c@{\extracolsep{0ptplus1fil}}
c@{\extracolsep{0ptplus1fil}}}\hline
$m_u$ & $m_s$ & $m_s/m_u$ & $M_0$ &   $M_u$ & $M_s$ & $M_s/M_u$  & $\kappa_0^{1/3}$ & $\kappa_\pi^{1/3}$ & $\kappa_K^{1/3}$ \\\hline
0.007  & 0.17 & 24.3 & 0.36 & 0.37 & 0.53 & 1.43  & 0.241 & 0.243 & 0.246
\\\hline
\end{tabular*}
\end{center}
\end{table}

In Table~\ref{Tab:DressedQuarks}, for reference, we report $T=0$ values of $u$- and $s$-quark properties, computed from Eq.\,\eqref{gapactual}.  The ratio $m_s/\bar m$, where $\bar m = (m_u+m_d)/2$, is consistent with contemporary estimates \cite{Leutwyler:2009jg}.  The result $M_s-m_s \approx M_0$ is typical \cite{Maris:1997tm,Bhagwat:2007ha} and indicates that the additive impact of DCSB is nearly as great for the $s$-quark as it is for $u,d$-quarks.  In general, however, $M_f-m_f$ is a monotonically decreasing function of $m_f$, bounded below by zero as $m_f\to\infty$ \cite{Bhagwat:2007ha,Holl:2005st}.

We have simplified the form of Eq.\,\eqref{gapactual} by introducing the function
\begin{eqnarray}
\nonumber
{\cal C}^{\rm iu}(\varsigma;T) & = &
8 T\! \sum_{l=-\infty}^\infty
\int_{\tau_{\rm uv}^2}^{\tau_{\rm ir}^2}\! d\tau \, {\rm e}^{-\tau(\varsigma+\omega_l^2)}
\int_0^\infty \! dq\, q^2 \, {\rm e}^{-\tau q^2}\\
&=& 2 T\! \sum_{l=-\infty}^\infty
\int_{\tau_{\rm uv}^2}^{\tau_{\rm ir}^2}\! d\tau \, {\rm e}^{-\tau(\varsigma+\omega_l^2)}
\frac{\surd \pi}{\tau^{3/2}}\\
&=& \int_{\tau_{\rm uv}^2}^{\tau_{\rm ir}^2}\! d\tau \, {\rm e}^{-\tau\varsigma}\,
2 T \vartheta_2({\rm e}^{-\tau 4 \pi^2 T^2} )\frac{\surd \pi}{\tau^{3/2}} \,,
\label{CirT}
\end{eqnarray}
where $\vartheta_2(x)$ is a Jacobi theta-function \cite{Gradshteyn:1980}.  It is straightforward to make the connection with zero temperature results once one appreciates that
\begin{equation}
\label{eq:2TJtheta0}
2 T \vartheta_2({\rm e}^{-\tau 4 \pi^2 T^2} ) \stackrel{T\to 0}{=} \frac{1}{\sqrt{\pi\tau}}\,.
\end{equation}

In deriving Bethe-Salpeter equations for vertices and bound-state amplitudes, the Ward-Green-Takahashi identities are crucial.  At $T\neq 0$, the identity in Eq.\,\eqref{WGTIT} is necessary and sufficient to ensure they are satisfied, where
\begin{eqnarray}
\nonumber
{\cal R}^{\rm iu}(\varsigma;T)
&=& \int_{\tau_{\rm uv}^2}^{\tau_{\rm ir}^2}\! d\tau \, {\rm e}^{-\tau\varsigma}\,
\sqrt{\frac{\pi}{\tau}}\\
&& \times \left[ -\frac{d}{d\tau}-\frac{1}{2}\frac{1}{\tau}\right]
2 T \vartheta_2({\rm e}^{-\tau 4 \pi^2 T^2} )\,.
\label{eq:RsT}
\end{eqnarray}
Using Eq.\,\eqref{eq:2TJtheta0}, it is straightforward to show that ${\cal R}^{\rm iu}(\varsigma;T\to0)=0$.

\section{Faddeev equations for $\Delta$-baryon}
\label{app:DeltaFE}
Here we explain the origin of our simple Faddeev equations for the baryons' dressed-quark-cores. 

Using a symmetry-preserving treatment of the contact interaction at zero temperature, the $\Delta$-baryon Faddeev amplitude can be written
\begin{equation}
\label{defpsiDelta}
\psi^\Delta_\mu(\ell,Q) = f^\Delta(\ell;Q) u^\Delta_\mu(Q)\,,
\end{equation}
where $u^\Delta_\mu(Q)$ is a Rarita-Schwinger spinor, defined in Eq.\,\eqref{rarita} of App.\,\ref{App:EM}.  (The reason for this simplicity is elucidated below.)  The amplitude is obtained from the following Faddeev equation:
\begin{eqnarray}
\nonumber
\lefteqn{f^\Delta(\ell_0;Q)u^\Delta_\mu(Q)}\\
& = & 4\int\frac{d^4\ell}{(2\pi)^4}
\mathcal M_{\mu\nu}^\Delta(\ell_0,\ell;Q)
f^\Delta(\ell;Q)u^\Delta_\nu(Q),
\end{eqnarray}
with $K_0=-\ell_0+Q$, $K=-\ell+Q$, $Q^2 = -m^2_\Delta$ and
\begin{eqnarray}
\nonumber \lefteqn{\mathcal M^\Delta_{\mu\nu}(\ell_0,\ell;P)}\\
& = &
 i\Gamma^{1^+}_\rho(K) S^{\rm T}(-\ell_0+K)
i{\bar \Gamma}^{1^+}_\mu(-K_0)S(\ell)\Delta^{1+}_{\rho\nu}(K)\,,\quad
\label{defCalM}
\end{eqnarray}
where
\begin{equation}
\label{propAVqq}
\Delta^{1+}_{\rho\nu}(K) = \frac{T_{\rho\nu}(K) }{K^2+m_{1^+}^2}\,,
T_{\rho\nu}(K) = \delta_{\rho\nu} + \frac{K_\rho K_\nu}{m_{1^+}^2}\,,
\end{equation}
is the axial-vector diquark's propagator and $\Gamma^{1^+}_\mu(K)$ is its Bethe-Salpeter amplitude:
\begin{equation}
\Gamma^{1^+}_\mu(K)C^\dagger = \gamma_\mu E_{1^+}(K).
\end{equation}

At this point, one post-multiplies by $\bar u_\beta(Q;r)$ and sums over the polarisation index to obtain [Eq.\,\eqref{Deltacomplete}],
\begin{equation}
\Lambda_+(Q) R_{\mu\beta}(Q)  = 4 \int\frac{d^4\ell}{(2\pi)^4}\,{\cal M}^\Delta_{\mu\nu}(\ell_0,\ell;Q)\, \Lambda_+(Q) R_{\nu\beta}(Q) \,,
\end{equation}
which, after contracting with $\delta_{\mu\beta}$, yields
\begin{eqnarray}
1 & = & 2{\rm tr}_{\rm D}\int \frac{d^4\ell}{(2\pi)^4}
\mathcal M_{\mu\nu}^\Delta(\ell_0,\ell;Q)
\Lambda_+(Q) R_{\nu\mu}(Q)\nonumber\\
& = & 2 E^2_{1^+}  {\rm tr}_{\rm D}\int \frac{d^4\ell}{(2\pi)^4}
\frac{\gamma_{\rho}[-i\gamma\cdot(\ell_0-{K})+M]\gamma_{\mu}}
{[K^2+m^2_{1^+}][(\ell_0-K)^2+M^2]}\nonumber\\
&& \frac{[-i\gamma\cdot\ell+ M] }{\ell^2+M^2}
\frac{[-i\gamma\cdot Q+m_\Delta]}{2 m_\Delta}  T_{\rho\nu}(K) R_{\nu\mu}(Q)\,. \label{eq:faddeev-delta-contracted}
\end{eqnarray}
(N.B.\ Here and below we suppress $f(\ell_0,Q)$, $f(\ell,Q)$ on the left- and right-hand sides, respectively, because, subject to our approximations, they will finally cancel.)

Previous $T=0$ studies of Eq.\,\eqref{eq:faddeev-delta-contracted} have used variants of the so-called ``static approximation,'' in which the quark exchanged in Fig.\,\ref{fig:Faddeev}, described by $S^{\rm T}(-\ell_0+K)$ in Eq.\,\eqref{defCalM}, is replaced by $g_\Delta^2/M$.  This expedient is discussed extensively in Sec.\,4 of Ref.\,\cite{Roberts:2011cf} and Sec.\,3.1 of Ref.\,\cite{Chen:2012qr}.  In starting with Eq.\,\eqref{defpsiDelta}, we assumed implicitly that a truncation of this sort would eventually be made, for only then does $\psi_\mu(\ell;Q)$ take such a simple form.  Indeed, in combination with diquark correlations generated by Eq.\,\eqref{njlgluon}, whose Bethe-Salpeter amplitudes are momentum-independent, the static approximation generates Faddeev equation kernels which are themselves momentum-independent and hence so are the Faddeev amplitudes.  The consequent simplifications are the merit of the truncation.

Unfortunately, the static approximation is inadequate at nonzero temperature because, in the chiral limit, the dressed-quark mass is expected to vanish for $T>T_c$.  In order to maintain context with the large body of work that has used the contact interaction \cite{GutierrezGuerrero:2010md,Roberts:2010rn,Roberts:2011cf,Roberts:2011wy,%
Wilson:2011aa,Chen:2012qr,Chen:2012tx}, we must provide a reasonable alternative.

To this end, consider that the right-hand-side of Eq.\,\eqref{eq:faddeev-delta-contracted} has the form
\begin{equation}
\int \frac{d^4\ell}{(2\pi)^4}
\frac{N(\ell,\ell-K_0,K,Q)}{[\ell^2+M^2][(\ell -K_0)^2+M^2][(-\ell+Q)^2+m_{1^+}^2]}\,,
\label{step1}
\end{equation}
where $N(\ell,\ell-K_0,K,Q)$ is a numerator and we have used the relation $(-\ell_0+K) = (-\ell + K_0)$.  In proceeding from this point, one employs a two-variable ($\alpha$, $\beta$) Feynman parametrisation to convert Eq.\,\eqref{step1} into
\begin{eqnarray}
\nonumber
\lefteqn{2 \int_0^1\!d\alpha \,d\beta\,\alpha\,}\\
&& \times \int \frac{d^4 l}{(2\pi)^4}
\frac{N(l+\eta_\mathpzc{p} Q, l- (\mathpzc{p}-\eta_\mathpzc{p}) Q,K,Q)}
{[l^2 + \varsigma_B(M^2,m_{1^+}^2,\alpha,\beta,\mathpzc{p},Q^2)]^3}\,,
\label{FEkernel}
\end{eqnarray}
where we have written $K_0 = \mathpzc{p}\, Q$; i.e., explicated that the external diquark carries a fraction $\mathpzc{p}$ of the baryon's momentum, defined $\eta_\mathpzc{p} = \hat\alpha + \alpha \beta \mathpzc{p}$ and also
\begin{eqnarray}
\nonumber \lefteqn{\varsigma_B(M^2,m_{1^+}^2,\alpha,\beta,\mathpzc{p},Q^2)=\alpha M^2}\\
&&
+\, \hat \alpha m_{1^+}^2+ [(\hat\alpha +\alpha\beta \mathpzc{p}^2)-
(\hat\alpha+\alpha\beta \mathpzc{p})^2]Q^2.
\end{eqnarray}

Every term in the denominator is real, and hence it acts as a weight function whose maximum occurs when $l^2=0$ and
\begin{equation}
\label{defineeta}
\eta_\mathpzc{p} = \hat\alpha + \alpha \beta \mathpzc{p} = \frac{1}{2} + \frac{m_{1^+}^2 - M^2}{2 Q^2}\,.
\end{equation}
We therefore define the integrand in the Faddeev equation as the quantity obtained from Eq.\,\eqref{FEkernel} with
\begin{equation}
\label{definemathpf}
\mathpzc{p} = 1-\eta_\mathpzc{p} = \frac{1}{2} + \frac{M^2-m_{1^+}^2}{2 Q^2}=: \mathpzc{p}_{1^+}\,.
\end{equation}
In so doing, we solve for the baryon Faddeev amplitude evaluated at a single value of the quark and diquark momenta; viz., respectively, $\ell_0 = (1-\mathpzc{p}_{1^+}) Q$, $K_0=\mathpzc{p}_{1^+} Q$.  (N.B.\ In a weak binding approximation, $m_{1^+}^2=4 M^2$, $Q^2=-m_B^2=-9M^2$ and hence $\mathpzc{p}_{1^+} = 2/3$.)

To be explicit, applying these rules in the present case maps Eq.\,\eqref{eq:faddeev-delta-contracted} into the following Faddeev equation for the $\Delta$-baryon:
\begin{eqnarray}
\nonumber 1 & = & \frac{E^2_{1^+}}{m_\Delta}
\int_0^1 \! d\alpha \int_0^1\!d\beta \, \alpha\, \\
&& \nonumber
\int \frac{d^4l}{(2\pi)^4}
\frac{1}
{[(l^2 + \varsigma_B(M^2,m_{1^+}^2,\alpha,\beta,\mathpzc{p}_{1^+},Q^2)]^3}\\
&& \nonumber
\times {\rm tr}_{\rm D}\bigg[ \gamma_{\rho}[-i\gamma\cdot(l + \eta Q -{K}_0)+M]\gamma_{\mu}[-i\gamma\cdot (l+\eta Q)\\
&& \quad    + M ][-i\gamma\cdot Q+m_\Delta]
T_{\rho\nu}(K) R_{\nu\mu}(Q)\bigg]\,.
\label{mappedDeltaFE}
\end{eqnarray}
The numerator in Eq.\,\eqref{mappedDeltaFE}, and analogous numerators in the nucleon Faddeev equation, produce many inner products.  We treat them as at $T=0$; viz., using the rules of Ref.\,\cite{Roberts:2011cf}:
\begin{subequations}
\begin{eqnarray}
&& (X\cdot l)(Y\cdot l) \to \frac{1}{4} \, l^2 X\cdot Y ,\; X,Y=Q,K,K_0, \quad \\
&& l\cdot Q \rightarrow \eta Q^2,\;
l\cdot K \rightarrow  \eta \mathpzc{p} \,Q^2, \;
l\cdot K_0 \rightarrow \eta \mathpzc{p}_f Q^2, \quad\\
&& K\cdot P\rightarrow \mathpzc{p}\, Q^2,\;
K_0\cdot P \rightarrow \mathpzc{p}_f Q^2,\\
&&  K_{0^+}^2 = -m_{0^+}^2 , \;  K_{1^+}^2 = -m_{1^+}^2,
\end{eqnarray}
\end{subequations}
with $\mathpzc{p} = 1-\eta$, $\eta = \hat\alpha+\alpha\beta \mathpzc{p}_f$; and $\mathpzc{p}_f = \mathpzc{p}_{1^+}$ in Eq.\,\eqref{definemathpf}, if the external diquark is an axial-vector, or
\begin{equation}
\label{definemathpf0}
\mathpzc{p}_f = \frac{1}{2} + \frac{M^2-m_{0^+}^2}{2 Q^2}=: \mathpzc{p}_{0^+}\,,
\end{equation}
if the external diquark is a scalar correlation.

We now define the nonzero temperature Faddeev equations by replacing the $dq_4$-integration in the expressions that arise by a Matsubara sum, so that
\begin{eqnarray}
\int\frac{d^4 l}{\pi^2}  \frac{1}{[l^2+\varsigma]^3} & \to & \overline{\cal C}_2^{\rm ir}(\varsigma;T)\,,\\
\int \frac{d^4 l}{\pi^2}   \frac{l^2}{[l^2+\varsigma]^3} &\to &
\big[ \overline{\cal C}_1^{\rm ir}(\varsigma;T) - \varsigma\overline{\cal C}_2^{\rm ir}(\varsigma;T)\big]\,,
\end{eqnarray}
where
\begin{equation}
\overline{\cal C}_1^{\rm ir}(\varsigma;T) = -\frac{d}{d\varsigma}\overline{\cal C}^{\rm ir}(\varsigma;T)\,,\;
\overline{\cal C}_2^{\rm ir}(\varsigma;T) = \frac{1}{2}\frac{d^2}{d\varsigma^2}\overline{\cal C}^{\rm ir}(\varsigma;T)\,,
\end{equation}
with $\overline{\cal C}^{\rm ir}(\varsigma;T)$ defined in Eq.\,\eqref{CirT}.

The procedure described above, applied to Eq.\,\eqref{mappedDeltaFE}, yields the following Faddeev equation for the $\Delta$:
\begin{equation}
1 = \mathpzc{K}^\Delta(Q^2=-m_\Delta^2)\,,
\end{equation}
with, using $\varsigma = \varsigma_\Delta(M^2,m_{1^+}^2,\alpha,\beta,\mathpzc{p}_{1^+},-m_\Delta^2)$,
\begin{eqnarray}
\nonumber
\lefteqn{ \mathpzc{K}^\Delta(Q^2=-m_\Delta^2)}\\
\nonumber
&=& \frac{E_{1^+}^2}{2\pi^2}
\int_0^1 \! d\alpha \, d\beta\, \alpha\,
\left\{
\overline{\mathcal C}_1^{\rm iu}(\varsigma;T)
\bigg[1+\frac{ m_\Delta ^2(\mathpzc{p}_{1^+}^2+ \mathpzc{p}^2)}{2 m_{1^+}^2}\bigg]\right.,\\
\nonumber
&& \overline{\mathcal C}_2^{\rm iu}(\varsigma;T)
\left[
-\frac{\varsigma}{2m_{1^+}^2} \right.
\left[ 2 m_{1^+}^2 + m_\Delta^2(\mathpzc{p}_{1^+}^2+\mathpzc{p}^2)\right]\\
\nonumber
&& + \frac{M+m_\Delta \eta}{3} \bigg(
6M + m_\Delta(2 \mathpzc{p} + 8 \mathpzc{p}_{1^+} - 6\eta)  \\
\nonumber
&& + \frac{m_\Delta^2}{m_{1^+}^2}
[3 M (\mathpzc{p}_{1^+}^2+\mathpzc{p}^2) + m_\Delta (4 \mathpzc{p} \, \mathpzc{p}_{1^+} \eta - 3 \mathpzc{p}_{1^+}^2 \eta+\\
&&
\left. \left. 2 \mathpzc{p}_{1^+} \mathpzc{p}^2-3\eta \mathpzc{p}^2)]
-\frac{4 \mathpzc{p}^2 \mathpzc{p}_{1^+}^2 \eta m_\Delta^5}{m_{1^+}^4}\bigg)\right]
\right\}. \label{FinalDeltaFaddeev}
\end{eqnarray}
Naturally, $E_{1^+}=E_{1^+}(T)$; i.e., the axial-vector diquark's Bethe-Salpeter amplitude is $T$-dependent, as are $M$, $m_{1^+}$.

At $T=0$, using the definitions in Table~\ref{Tab:DressedQuarks}, this Faddeev equation produces
\begin{equation}
\label{Deltamass}
m_\Delta(T=0) = 1.42\,{\rm GeV}\,.
\end{equation}
Equation~\eqref{FinalDeltaFaddeev} represents an improvement over previous definitions of the static approximation because it eliminates one parameter yet produces the same mass (within 2\%) as that fitted using the extra parameter.

We reiterate here that this equation describes the $\Delta$-baryon's ``dressed-quark-core''.  It should and does, therefore, produce a mass that lies above that quoted empirically for the $\Delta$.  Remarkably, indeed, the value produced is almost identical to that inferred for the dressed-quark core via the dynamical coupled-channels analysis of Ref.\,\cite{Suzuki:2009nj}.  Any theoretical framework that produces a stable (width zero) $\Delta$-resonance with a mass near that quoted empirically is dubious.  In this connection it is also noteworthy that $m_\Delta/m_\rho = 1.53$, a result which compares favourably with the experimental value: 1.59.

\section{Nucleon Faddeev equation}
\label{app:Nucleon}
We capitalise on Eq.\,(C.47) in Ref.\,\cite{Roberts:2011cf} to write the nucleon Faddeev equation as a $3\times 3$ matrix eigenvalue problem for the scalar $[ud]$ and axial-vector $\{ud\}$ diquark correlations.  To begin, the nucleon is described by a Faddeev amplitude:
\begin{equation}
\label{psiN}
\psi_{\{\mu\}}^N(Q)u(Q) =
\left[\begin{array}{l}
\mathpzc{s}(Q) \\
\mathpzc{a}_\mu(Q)
\end{array}\right]u(Q)\,,
\end{equation}
where
\begin{subequations}
\label{psiNdetail}
\begin{eqnarray}
\mathpzc{s}(Q) & = & s(Q) \mathbf{I}_{\rm D} \,,\\
\mathpzc{a}_\mu(Q)& = & a_1(Q) \gamma_5\gamma_\mu + a_{2}(Q)\gamma_5 \hat Q_\mu \,,
\end{eqnarray}
\end{subequations}
with $\hat Q^2=-1$ and $u(Q)$ being the spinor introduced in Eq.\,\eqref{nucleonspinor}.

The amplitude satisfies
\begin{eqnarray}
\nonumber \lefteqn{\psi_{\{\mu\}}^N(Q)u(Q) }\\
&=& -4\int\frac{d^4\ell}{(2\pi)^4} {\cal M}_{\{\mu\nu\}}(\ell_0,\ell;Q)\psi_{\{\nu\}}^N(Q)u(Q)\,,
\end{eqnarray}
with
\begin{equation}
\label{calM} {\cal M}_{\{\mu\nu\}}(\ell_0,\ell;P) = \left[\begin{array}{cc}
{\cal M}_{00} & 3({\cal M}_{01})_\nu \\
({\cal M}_{10})_\mu & -({\cal M}_{11})_{\mu\nu}\rule{0mm}{3ex}
\end{array}
\right] ,
\end{equation}
where $(l_0^K=\ell_0-K)$
\begin{subequations}
\begin{eqnarray}
\nonumber
\lefteqn{{\cal M}_{00}}\\
& =& \Gamma^{0^+}\!(K)\,
S^{\rm T}(-\ell_0^K) \,\bar\Gamma^{0^+}\!(-K_0)\,
S(\ell)\,\Delta^{0^+}(K) \,,\label{calM00} \\
\nonumber \lefteqn{({\cal M}_{01})_\nu }\\
&=& \Gamma_\mu^{1^+}\!(K) S^{\rm T}(-\ell_{0}^K)\,
\bar\Gamma^{0^+}\!(-K_0)\,
S(\ell)\,\Delta^{1^+}_{\mu\nu}(K) , \rule{2.2em}{0ex} \label{calM01} \\
\nonumber
\lefteqn{({\cal M}_{10})_\mu }\\
&=& \Gamma^{0^+}\!(K)\, S^{\rm T}(-\ell_0^K)\, \bar\Gamma_\mu^{1^+}\!(-K_0)\,
S(\ell)\,\Delta^{0^+}(K) , \label{calM10} \\
\nonumber
\lefteqn{({\cal M}_{11})_{\mu\nu} }\\
&=&
\Gamma_\rho^{1^+}\!(K)\, S^{\rm T}(-\ell_{0}^K)\,
\bar\Gamma^{1^+}_\mu\!(-K_0)\,
S(\ell)\,\Delta^{1^+}_{\rho\nu}(K) \,. \label{calM11}
\end{eqnarray}
\end{subequations}
In Eqs.\,\eqref{calM00}--\eqref{calM10},
\begin{equation}
\label{propSCqq}
\Delta^{0+}(K) = \frac{1}{K^2+m_{0^+}^2}
\end{equation}
is the scalar diquark's propagator and $\Gamma_{0^+}(K)$ its Bethe-Salpeter amplitude:
\begin{equation}
\Gamma_{0^+}(K)C^\dagger
= i \gamma_5 E_{0^+}(K) + \frac{1}{M}\gamma_5 \gamma\cdot K F_{0^+}(K)\,.
\end{equation}

At this point one can follow the procedure in App.\,\ref{app:DeltaFE} and thereby arrive at the following eigenvalue problem for $Q^2=-m_N^2$:
\begin{equation}
\label{NFaddeev}
\psi(Q)=\left[
\begin{array}{c}
s(Q)\\
a_1(Q)\\
a_2(Q)
\end{array}\right]
= \mathcal K(Q)\psi(Q)
\end{equation}
where the kernel is a $3\times 3$ matrix:
\begin{equation}
\label{NucleonKernel}
\mathcal K(Q) = \left[
\begin{array}{ccc}
\mathcal K_{ss}^{00} & 3 \mathcal K_{sa_1}^{01} & 3 \mathcal K_{sa_2}^{01}\\
\mathcal K_{a_1s}^{10} & - \mathcal K_{a_1a_1}^{11} & - \mathcal K_{a_1a_2}^{11}\\
\mathcal K_{a_2s}^{10} & - \mathcal K_{a_2a_1}^{11} & - \mathcal K_{a_2a_2}^{11}
\end{array}
\right].
\end{equation}
The entries in Eq.\,\eqref{NucleonKernel} are described below.  N.B.\,The Bethe-Salpeter amplitudes, dressed-quark mass and diquark masses are $T$-dependent.

\begin{subequations}
{\allowdisplaybreaks
\begin{eqnarray}
\nonumber
K_{ss}^{00}&=&\frac{1}{2\pi^2}\left[ E_{0^+}^2 \mathcal K_{EE}+E_{0^+}F_{0^+} \mathcal K_{EF} \right . \\
&& \left. + F_{0^+}^2 \mathcal K_{FF} \right] ,\\
\nonumber \mathcal K_{EE}
& = & \int_0^1 \! d\alpha\, d\beta \, \alpha\,
\big\{
\overline{\mathcal C}_1^{\rm iu}(\varsigma;T)
+ \overline{\mathcal C}_2^{\rm iu}(\varsigma;T)
[( M  \\
&& \quad \quad + m_N [\mathpzc{p}_{0^+}-\eta ]) (M+m_N \eta )-\varsigma] \big\},\\
\nonumber \mathcal K_{EF}
& = &
\frac{m_N}{2 M}\int_0^1 \! d\alpha d\beta\, \alpha\,
\left\{
\overline{\mathcal C}_1^{\rm iu}(\varsigma;T)(\mathpzc p_{0^+}-2 \mathpzc p)
\right. \\
&&
\quad \quad - \overline{\mathcal C}_2^{\rm iu}(\varsigma;T)
[ 2(\mathpzc p_{0^+}+\mathpzc p)(M+m_N \eta ) (M \nonumber\\
&&
\left. \quad \quad + m_N (\mathpzc p_{0^+}-\eta )) + \varsigma(\mathpzc p_{0^+}-2\mathpzc p)] \rule{0em}{2.5ex}\right\},\\
\nonumber \mathcal K_{FF}
& = & -\frac{m_N^2 \mathpzc p_{0^+} }{2 M^2}
\int_0^1 \! d\alpha d\beta\, \alpha\,\mathpzc p\,
\left\{\rule{0em}{2.5ex} \overline{\mathcal C}_1^{\rm iu}(\varsigma;T)
-\overline{\mathcal C}_2^{\rm iu}(\varsigma;T) [\varsigma  \right.\\
&& \left. \rule{0em}{2.5ex}
+ 2 (M + m_N (\mathpzc p_{0^+}-\eta )) (M+m_N \eta ) ]\right\},
\end{eqnarray}}
\end{subequations}
\hspace*{-0.4\parindent}where $\varsigma = \varsigma_B(M^2,m_{0^+}^2,\alpha,\beta,\mathpzc{p}_{0^+},-m_N^2)$,
$\mathpzc p = 1- \eta$,
$\eta = \hat\alpha + \alpha\beta \mathpzc{p}_{0^+}$.

\begin{subequations}
{\allowdisplaybreaks
\begin{eqnarray}
\mathcal K_{\;sa_1}^{01}&=& \frac{E_{1^+}}{2\pi^2}[ E_{0^+} \mathcal K_{E a_1}+F_{0^+} \mathcal K_{F a_1} ],\\
\nonumber
\mathcal K_{E a_1} & = & \int_0^1 \! d\alpha d\beta\, \alpha\,
\left\{\rule{0em}{2.5ex}
 3\overline{\mathcal C}_1^{\rm iu}(\varsigma;T)
+ \overline{\mathcal C}_2^{\rm iu}(\varsigma;T) \bigg[3 M^2 - 3 \varsigma   \right. \\
&& \nonumber
\quad + \; 3 m_N^2 \eta(\mathpzc p_{0^+}-\eta) + \mathpzc p_{0^+} M m_N \\
&& \left.
\quad + \frac{2 \mathpzc p^2 \mathpzc p_{0^+} M m_N^3}{m_{1^+}^2} \bigg]
\rule{0em}{2.5ex} \right\},\\
\nonumber
\mathcal K_{F a_1} &=& \frac{m_N \mathpzc p_{0^+}}{2M}
\int_0^1 \! d\alpha d\beta\, \alpha\,
\left\{\rule{0em}{2.5ex}
\overline{\mathcal C}_1^{\rm iu}(\varsigma;T)
\left[1+\frac{ 2 m_N^2 p^2}{m_{1^+}^2}\right] \right.\\
&& \nonumber
- \overline{\mathcal C}_2^{iu}(\varsigma;T)
\bigg[2 M^2 +6 M m_N \mathpzc p_{0^+} + \varsigma \\
&& \nonumber
\quad +\frac{2 \mathpzc p^2  m_N^2}{ m_{1^+}^2} (2 M^2 +2 m_N^2  \eta(\mathpzc p_{0^+}-\eta) + \varsigma ) \\
&&
\quad + 2 m_N^2 (\mathpzc p_{0^+}-\eta ) \eta
\left. \bigg] \rule{0em}{2.5ex} \right\};
\end{eqnarray}}
\end{subequations}
\hspace*{-0.4\parindent}and
\begin{subequations}
{\allowdisplaybreaks
\begin{eqnarray}
\mathcal K_{\;sa_2}^{01} &=& \frac{E_{1^+}}{2\pi^2}[ E_{0^+} \mathcal K_{E a_2}+F_{0^+} \mathcal K_{F a_2} ],\\
\nonumber
\mathcal K_{E a_2} & = &
\int_0^1 \! d\alpha d\beta\, \alpha\,\mathpzc r_{1^+}\,
\left\{\rule{0em}{2.5ex} \overline{\mathcal C}_1^{\rm iu}(\varsigma;T)
- \overline{\mathcal C}_2^{\rm iu}(\varsigma;T) [\varsigma
\right. \\
&& +(M-m_N\eta)(m_N(\mathpzc p_{0^+}-\eta)-M)] \left. \rule{0em}{2.5ex} \right\}, \quad \quad \\
\nonumber
\mathcal K_{F a_2} & = &
-\frac{\mathpzc p_{0^+} m_N}{2M} \int_0^1 \! d\alpha d\beta\, \alpha\,\mathpzc r_{1^+}\,
\left\{\rule{0em}{2.5ex}
\overline{\mathcal C}_1^{\rm iu}(\varsigma;T)
- \overline{\mathcal C}_2^{\rm iu}(\varsigma;T)
[ \varsigma \right. \\
&&  + 2 (M - m_N \eta ) (M+m_N [\eta - \mathpzc p_{0^+}])]
\left. \rule{0em}{2.5ex} \right\},
\end{eqnarray}}
\end{subequations}
\hspace*{-0.4\parindent}where $\varsigma = \varsigma_B(M^2,m_{1^+}^2,\alpha,\beta,\mathpzc{p}_{0^+},-m_N^2)$, $\mathpzc r_{1^+} = 1 - m_N^2 \mathpzc p^2/m_{1^+}^2 $, $\mathpzc p = 1- \eta$, $\eta = \hat\alpha + \alpha\beta \mathpzc{p}_{0^+}$.

\begin{subequations}
{\allowdisplaybreaks
\begin{eqnarray}
\mathcal K_{a_1 s}^{10}&=& \frac{E_{1^+}}{2\pi^2}[ E_{0^+} \mathcal K_{\;a_1 E}+F_{0^+} \mathcal K_{\;a_1 F} ],\\
\nonumber
\mathcal K_{\;a_1 E} &=&
\int_0^1 \! d\alpha d\beta\, \alpha\,
\left\{\rule{0em}{2.5ex}
\frac{2 m_{1^+}^2 + m_N^2 \mathpzc p_{1^+}^2}{6 m_{1^+}^2}
[\overline{\mathcal C}_1^{\rm iu}(\varsigma;T) \quad
\right.\\
&& \nonumber
\quad + \overline{\mathcal C}_2^{\rm iu}(\varsigma;T) (2 M^2 -2 m_N^2 \eta^2- \varsigma ) ]\\
&&
\left. \quad + \overline{\mathcal C}_2^{\rm iu}(\varsigma;T)
\mathpzc p_{1^+} m_N (M+\eta m_N) \rule{0em}{3ex}  \right\},\\
%
%
\nonumber
\mathcal K^{10}_{\;a_1 F} &=& - \frac{ m_N  }{M }
\int_0^1 \! d\alpha d\beta\, \alpha\,\mathpzc p\,
\left\{\rule{0em}{2.5ex}
\frac{2 m_{1^+}^2 +m_N^2 \mathpzc p_{1^+}^2}{6 m_{1^+}^2}
[ \overline{\mathcal C}_1^{\rm iu}(\varsigma;T)
\right. \\
&& + \overline{\mathcal C}_2^{iu}(\varsigma;T) (2 M^2 - 2 m_N^2 \eta ^2-\varsigma)]\\
&& \left. + \overline{\mathcal C}_2^{iu}(\varsigma;T) \mathpzc p_{1^+} m_N  (M+ \eta m_N)
\rule{0em}{3ex}  \right\}
\end{eqnarray}}
\end{subequations}
\hspace*{-0.4\parindent}where $\varsigma = \varsigma_B(M^2,m_{0^+}^2,\alpha,\beta,\mathpzc{p}_{1^+},-m_N^2)$,
$\mathpzc p = 1- \eta$, $\eta = \hat\alpha + \alpha\beta \mathpzc{p}_{1^+}$.

\begin{subequations}
{\allowdisplaybreaks
\begin{eqnarray}
\mathcal K_{a_2s}^{10}&=& \frac{E_{1^+}}{2\pi^2}[E_{0^+} \mathcal K_{\; a_2 E}+F_{0^+} \mathcal K_{\;a_2F}]\\
\nonumber
\mathcal K_{\;a_2 E} &=& \int_0^1 \! d\alpha d\beta\, \alpha\,
\left\{ \rule{0em}{2.5ex}
\frac{ m_{1^+}^2 - 4 m_N^2 \mathpzc p_{1^+}^2}{6 m_{1^+}^2} [
\overline{\mathcal C}_1^{\rm iu}(\varsigma;T)
\right. \\
&& \nonumber
- \varsigma \overline{\mathcal C}_2^{\rm iu}(\varsigma;T)]
+ \frac{1}{3}\overline{\mathcal C}_2^{\rm iu}(\varsigma;T) (M+\eta m_N)\\
&&
\quad \times [M+m_N(5\eta-3p_{1^+})\nonumber\\
&& \quad
-\frac{2 p_{1^+}^2m_N^2}{m_{1^+}^2}(2M+\eta m_N)]
\left.  \rule{0em}{2.5ex} \right\},\\
\nonumber
\mathcal K_{\;a_2 F} &=& \frac{m_N \mathpzc p}{6 M } \int_0^1 \! d\alpha d\beta\, \alpha\,
\left\{ \rule{0em}{2.5ex}
\frac{5m_{1^+}^2 - 2 m_N^2 p_{1^+}^2}{m_{1^+}^2}
[\overline{\mathcal C}_1^{\rm iu}(\varsigma;T)
\right.\\
\nonumber
&& + \overline{\mathcal C}_2^{\rm iu}(\varsigma;T) (2 M^2 -\varsigma) ]
+ \overline{\mathcal C}_2^{\rm iu}(\varsigma;T) [2\eta m_N^2(3p_{1^+} \\
&& \nonumber
\quad + \eta) +6M m_N(\mathpzc{p}_{1^+}+2\eta)-\frac{2  m_N^2 p_{1^+}^2}{m_{1^+}^2}\\
&& \quad \times (4 \eta^2 m_N^2+6\eta M m_N]
 \left. \rule{0em}{2.5ex} \right\},
\end{eqnarray}}
\end{subequations}
\hspace*{-0.4\parindent}where $\varsigma = \varsigma_B(M^2,m_{0^+}^2,\alpha,\beta,\mathpzc{p}_{1^+},-m_N^2)$,
$\mathpzc p = 1- \eta$, $\eta = \hat\alpha + \alpha\beta \mathpzc{p}_{1^+}$.

{\allowdisplaybreaks
\begin{eqnarray}
\nonumber
\mathcal K_{\; a_1a_1}^{11} & = & - \frac{E_{1^+}^2}{6\pi^2}
\int_0^1 \! d\alpha d\beta\, \alpha\,
\left\{ \rule{0em}{2.5ex}
\overline{\mathcal C}_1^{\rm iu}(\varsigma;T)
\left[1+\frac{m_N^2}{2 m_{1^+}^2} \right.
\left(\mathpzc p_{1^+}^2 \right.
\right. \\
\nonumber
&&  \left. + 2 \mathpzc p \mathpzc p_{1^+}-2\mathpzc p^2)\rule{0em}{3ex}\right]
-\overline{\mathcal C}_2^{\rm iu}(\varsigma;T)
\left[\rule{0em}{3ex}  M  m_N (\mathpzc p_{1^+}-4\mathpzc p) \right. \\
&& + m_N^2 \eta  (2 \eta -5 p_{1^+})-2M^2+\varsigma\nonumber\\
&& + \frac{m_N^2}{2m_{1^+}^2} \big[  (\mathpzc p_{1^+}^2-2\mathpzc p^2+2 \mathpzc p_{1^+} \mathpzc p)\varsigma+2 M^2(2\mathpzc p^2 \nonumber\\
&&-\mathpzc p_{1^+}^2-2\mathpzc p_{1^+} \mathpzc p)+ 2 \eta m_N^2   (2 \mathpzc p_{1^+} \mathpzc p^2-2\eta \mathpzc p^2+ \nonumber\\
&&
\left. \left.
\mathpzc p_{1^+}^2\eta +6\eta \mathpzc p_{1^+} \mathpzc p  )  \big]
-\frac{ 4m_N^6 \mathpzc p_{1^+}^2 \mathpzc p^2\eta^2}{m_{1^+}^4} \right] \rule{0em}{3ex} \right\},\\
\nonumber
\mathcal K_{\; a_1a_2}^{11}&=& - \frac{E_{1^+}^2}{6\pi^2}
\int_0^1 \! d\alpha d\beta\, \alpha\,
\left\{\rule{0em}{4ex}
\overline{\mathcal C}_1^{\rm iu}(\varsigma;T)\left[\rule{0em}{2.5ex}\right.
1+\frac{m_N^2}{2 m_{1^+}^2}( \mathpzc p_{1^+}^2 \right. \\
\nonumber
&&
- \mathpzc p \mathpzc p_{1^+} -2 \mathpzc p^2)\left.\rule{0em}{2.5ex}\right]
-\overline{\mathcal C}_2^{\rm iu}(\varsigma;T)
\left( \rule{0em}{4ex}
\varsigma \left[\rule{0em}{2.5ex} \right.
1 +\frac{m_N^2}{2m_{1^+}^2}(\mathpzc p_{1^+}^2 \right.\\
\nonumber
&& -\mathpzc p_{1^+} \mathpzc p -2\mathpzc p^2)\left. \rule{0em}{2.5ex} \right]
+(\eta m_N-M) \left[\rule{0em}{4ex} \right. (2\mathpzc p-3\mathpzc p_{1^+}\\
\nonumber
&& +2\eta) m_N+ 2 M+ \frac{m_N^2}{m_{1^+}^2}
    (M [\mathpzc p_{1^+}^2-\mathpzc p_{1^+} \mathpzc p-2 \mathpzc p^2]\\
\nonumber
&&
+m_N [ \mathpzc p_{1^+}^2\eta+\mathpzc p \mathpzc p_{1^+} \eta +
\mathpzc p^2 (\mathpzc p_{1^+} -2 \eta) ])\\
&&
\left. \left. \left.
-\frac{2 \mathpzc p^2 \mathpzc p_{1^+}^2\eta m_N^5}{m_{1^+}^4}\right]
\rule{0em}{2.5ex}\right) \rule{0em}{3ex}\right\}, \\
\nonumber
\mathcal K_{a_2a_1}^{11}&=& - \frac{E_{1^+}^2}{12\pi^2}
\int_0^1 \! d\alpha d\beta\, \alpha\,
\left\{ \rule{0em}{4ex}
\overline{\mathcal C}_1^{\rm iu}(\varsigma;T)
\left[1-\frac{4 m_N^2}{m_{1^+}^2}( \mathpzc p_{1^+}^2
\right.\right. \\
&& \left.
+2 \mathpzc p \mathpzc p_{1^+}-2\mathpzc p^2)\rule{0em}{3ex}\right]
-\overline{\mathcal C}_2^{\rm iu}(\varsigma;T)
\left(\rule{0em}{3ex}
\varsigma -2M^2\nonumber
\right. \\
&& + 2M m_N(4 \mathpzc p- \mathpzc p_{1^+}-18\eta) + 2 \eta m_N^2(5 \mathpzc p_{1^+}\nonumber\\
&& + 6 \mathpzc p-5 \eta) + \frac{1}{m_{1^+}^2}[4m_N^2(\mathpzc p_{1^+} ^2-2 \mathpzc p^2+2 \mathpzc p \mathpzc p_{1^+}) \nonumber\\
&&\times (2M^2-\varsigma)+36 \mathpzc p_{1^+}^2\eta M m_N^3+4 \eta m_N^4\nonumber\\
&&( \eta [ \mathpzc p_{1^+}^2 - 6\mathpzc p_{1^+} \mathpzc p -2\mathpzc p^2] -4\mathpzc p_{1^+} \mathpzc p^2 )]\nonumber\\
&&
\left.\left.
+\frac{32 \mathpzc p_{1^+}^2 \mathpzc p^2\eta^2m_N^6}{m_{1^+}^4}
\right) \rule{0em}{4ex} \right\}, \\
\nonumber
\mathcal K_{\;a_2a_2}^{11}&=& \frac{E_{1^+}^2}{12\pi^2}
\int_0^1 \! d\alpha d\beta\, \alpha\,
\left\{ \rule{0em}{4ex}
\overline{\mathcal C}_1^{\rm iu}(\varsigma;T)
\left[ 5  +\frac{m_N^2}{ m_{1^+}^2}(2 \mathpzc p_{1^+} p
\right.\right.\\
&& \nonumber
\left. -2 \mathpzc p_{1^+}^2 -5 \mathpzc p^2)\rule{0em}{3.5ex} \right]
+ \overline{\mathcal C}_2^{\rm iu}(\varsigma;T)
\left(\rule{0em}{3ex} \varsigma \left[\frac{m_N^2}{m_{1^+}^2}
(2\mathpzc p_{1^+}^2  \right.
\right. \\
&&
\left.\rule{0em}{3ex} +5 \mathpzc p^2 -2 \mathpzc p_{1^+} \mathpzc p)-5\right]
+(M-m_N\eta) \bigg[10M\nonumber\\
&&-2m_N(3 \mathpzc p_{1^+}+\eta+\mathpzc p)+\frac{2m_N^2}{m_{1^+}^2}(M[2 \mathpzc p_{1^+} \mathpzc p-5\mathpzc p^2\nonumber\\
&&-2\mathpzc p_{1^+}^2] +m_N [\eta ( 4\mathpzc p_{1^+} \mathpzc p+ \mathpzc p^2+4 \mathpzc p_{1^+}^2) +4 \mathpzc p_{1^+} \mathpzc p^2 ])\nonumber\\
&&
\left. -\frac{16m_N^5\mathpzc p_{1^+}^2\mathpzc p^2\eta}{m_{1^+}^4}\bigg]
\rule{0em}{4ex} \right\},
\end{eqnarray}}
\hspace*{-0.4\parindent}where $\varsigma = \varsigma_B(M^2,m_{1^+}^2,\alpha,\beta,\mathpzc{p}_{1^+},-m_N^2)$,
$\mathpzc p = 1- \eta$, $\eta = \hat\alpha + \alpha\beta \mathpzc{p}_{1^+}$.

At $T=0$, using the definitions in Table~\ref{Tab:DressedQuarks}, the nucleon Faddeev equation yields
\begin{equation}
\label{MassN}
m_N(T=0) = 1.09\,{\rm GeV}\,,
\end{equation}
which is just 4\% less than the value produced by the static approximation employed in Ref.\,\cite{Chen:2012qr}.  The Faddeev amplitude is
\begin{equation}
\label{FaddeevAN}
\begin{array}{ccc}
s &a_1 &a_2 \\
0.85 & 0.41 &-0.34
\end{array},
\end{equation}
which may be compared with the static approximation result \cite{Chen:2012qr}: $(s,a_1,a_2) = (0.88,0.47,-0.078)$.  Plainly once again, Eqs.\,\eqref{NFaddeev}, \eqref{NucleonKernel} represent an improvement over previous definitions of the static approximation because they eliminate one parameter yet produce essentially the same mass.

Here, too, we emphasise that Eqs.\,\eqref{NFaddeev}, \eqref{NucleonKernel} describe the nucleon's ``dressed-quark-core''.  They should and do, therefore, produce a mass that lies above that quoted empirically.  The inclusion of resonant contributions to the kernel leads typically to a 0.15\,GeV reduction in the bound-state's mass \cite{Hecht:2002ej,Cloet:2008re}.  Notably, the mass in Eq.\,\eqref{MassN} is within 12\% of the undressed nucleon's mass in Ref.\,\cite{Gasparyan:2003fp}.  Moreover, $m_N/m_\rho = 1.17$, which compares favourably with the experimental value of $1.21$.

\section{Euclidean Conventions}
\label{App:EM}
Our $T=0$ Euclidean conventions are specified here.
\begin{equation}
p\cdot q=\sum_{i=1}^4 p_i q_i\,;
\end{equation}
\begin{eqnarray}
&& \{\gamma_\mu,\gamma_\nu\}=2\,\delta_{\mu\nu}\,;\;
\gamma_\mu^\dagger = \gamma_\mu\,;\;
\sigma_{\mu\nu}= \frac{i}{2}[\gamma_\mu,\gamma_\nu]\,; \rule{2em}{0ex}\\
&& {\rm tr}\,[\gamma_5\gamma_\mu\gamma_\nu\gamma_\rho\gamma_\sigma] =
-4\,\epsilon_{\mu\nu\rho\sigma}\,, \epsilon_{1234}= 1\,.
\end{eqnarray}

A positive energy spinor satisfies
\begin{equation}
\label{nucleonspinor}
\bar u(P,s)\, (i \gamma\cdot P + M) = 0 = (i\gamma\cdot P + M)\, u(P,s)\,,
\end{equation}
where $s=\pm$ is the spin label.  It is normalised:
\begin{equation}
\bar u(P,s) \, u(P,s) = 2 M \,,
\end{equation}
and may be expressed explicitly:
\begin{equation}
u(P,s) = \sqrt{M- i {\cal E}}
\left(
\begin{array}{l}
\chi_s\\
\displaystyle \frac{\vec{\sigma}\cdot \vec{P}}{M - i {\cal E}} \chi_s
\end{array}
\right)\,,
\end{equation}
with ${\cal E} = i \sqrt{\vec{P}^2 + M^2}$,
\begin{equation}
\chi_+ = \left( \begin{array}{c} 1 \\ 0  \end{array}\right)\,,\;
\chi_- = \left( \begin{array}{c} 0\\ 1  \end{array}\right)\,.
\end{equation}
For the free-particle spinor, $\bar u(P,s)= u(P,s)^\dagger \gamma_4$.

The spinor can be used to construct a positive energy projection operator:
\begin{equation}
\label{Lplus} \Lambda_+(P):= \frac{1}{2 M}\,\sum_{s=\pm} \, u(P,s) \, \bar
u(P,s) = \frac{1}{2M} \left( -i \gamma\cdot P + M\right).
\end{equation}

A negative energy spinor satisfies
\begin{equation}
\bar v(P,s)\,(i\gamma\cdot P - M) = 0 = (i\gamma\cdot P - M) \, v(P,s)\,,
\end{equation}
and possesses properties and satisfies constraints obtained via obvious analogy
with $u(P,s)$.

A charge-conjugated Bethe-Salpeter amplitude is obtained via
\begin{equation}
\label{chargec}
\bar\Gamma(k;P) = C^\dagger \, \Gamma(-k;P)^{\rm T}\,C\,,
\end{equation}
where ``T'' denotes a transposing of all matrix indices and
$C=\gamma_2\gamma_4$ is the charge conjugation matrix, $C^\dagger=-C$.  We note that
\begin{equation}
\label{chargecT}
C^\dagger \gamma_\mu^{\rm T} \, C = - \gamma_\mu\,, \; [C,\gamma_5] = 0\,.
\end{equation}

In describing the decuplet $\Delta$-baryon we employ a Rarita-Schwinger spinor to represent a covariant spin-$3/2$ field.  The positive energy
spinor is defined by the following equations:
\begin{subequations}
\label{rarita}
{\allowdisplaybreaks
\begin{eqnarray}
(i \gamma\cdot P + M)\, u_\mu(P;r) &=& 0\,,\; \\
\gamma_\mu u_\mu(P;r) &=& 0\,,\; \\
P_\mu u_\mu(P;r) &=& 0\,,
\end{eqnarray}}
\end{subequations}
\hspace*{-0.4\parindent}where $r=-3/2,-1/2,1/2,3/2$.  It is normalised:
\begin{equation}
\bar u_{\mu}(P;r^\prime) \, u_\mu(P;r) = 2 M\,,
\end{equation}
and satisfies a completeness relation
\begin{equation}
\label{Deltacomplete}
\frac{1}{2 M}\sum_{r=-3/2}^{3/2} u_\mu(P;r)\,\bar u_\nu(P;r) =
\Lambda_+(P)\,R_{\mu\nu}\,,
\end{equation}
where
\begin{equation}
R_{\mu\nu} = \delta_{\mu\nu} \mbox{\boldmath $I$}_{\rm D} -\frac{1}{3} \gamma_\mu \gamma_\nu +
\frac{2}{3} \hat P_\mu \hat P_\nu \mbox{\boldmath $I$}_{\rm D} - i\frac{1}{3} [ \hat P_\mu
\gamma_\nu - \hat P_\nu \gamma_\mu]\,,
\end{equation}
with $\hat P^2 = -1$, which is very useful in simplifying the Faddeev equation for a positive energy decuplet state.



\end{document}